\documentclass[12pt,preprint]{aastex}

\slugcomment{To appear in the Astronomical Journal, January 2005}

\shorttitle{VY CMa}
\shortauthors{Humphreys   et al.}

\begin{document}


\title{High Resolution, Long -- Slit Spectroscopy of VY CMa: The Evidence for Localized High Mass Loss Events
} 

\author{Roberta M. Humphreys, Kris Davidson, and Gerald Ruch}
\affil{Astronomy Department, University of Minnesota,
    Minneapolis, MN 55455}
\email{roberta@aps.umn.edu}

\and

\author{George Wallerstein}
\affil{Astronomy Department, University of Washington, Seattle, WA 98195}


\begin{abstract}
High spatial and spectral resolution spectroscopy of the OH/IR supergiant  
VY CMa and its circumstellar ejecta reveals evidence for high mass loss events from 
localized regions on the star occurring over the past 1000 years. 
The reflected absorption lines and the extremely
strong K I emission lines show a complex pattern of velocities in the ejecta.  
We show that the large, dusty NW arc,  expanding at $\approx$ 50 km s$^{-1}$ with 
respect to the embedded star, is kinematically distinct from the surrounding nebulosity 
and was ejected about 400 years ago. 
Other large, more filamentary loops were probably expelled as much as 800 to 1000 years ago while
knots and small arcs close to the star resulted from more recent events 100 to 200 years
ago.  {\it The more diffuse, uniformly 
distributed gas and dust is surprisingly stationary with little or no velocity 
relative to the star.} This is not what we would expect for the circumstellar material
from an evolved red supergiant with a long history of mass loss. We therefore suggest that the 
high mass loss rate for VY CMa is a measure of the mass  carried out by these specific 
ejections accompanied by streams or flows of gas through low density regions
in the dust envelope.  VY CMa may thus  be our most extreme example of stellar activity, but
our results also bring into question the evolutionary state of this famous star.  
In a separate Appendix, we discuss the origin of the very strong K I and other rare
emission lines in its spectrum.
\end{abstract}


%% subject header.  Objects should be in the appropriate "individual"
%% headers (e.g. quasars: individual, stars: individual, etc.) with the
%% additional provision that the total number of headers, including each
%% individual object, not exceed six.  The \objectname{} macro, and its
%% alias \object{}, is used to mark each object.  The macro takes the object
%% name as its primary argument.  This name will appear in the paper
%% and serve as the link's anchor in the electronic edition if the name
%% is recognized by the data centers.  The macro also takes an optional
%% argument in parentheses in cases where the data center identification
%% differs from what is to be printed in the paper.



\section{Introduction}
The powerful infrared source and OH maser VY Canis Majoris is one of 
a few very luminous cool hypergiant stars that define 
the empirical upper luminosity boundary in the 
HR diagram \citep{HD94,deJ98}. It  is  thought to represent a very rapid stage
of evolution characterized by a  high mass loss rate. The red supergiant stage
is a few hundred thousand years long, or less for the most massive stars,
but the transition time  from  main sequence to RSG and later from RSG back to a yellow
hypergiant may be only a few thousand years \citep{Sch92,Sch93}. These evolved late-type stars 
are  major sources of gas, dust and molecules that contribute 
to the replenishment of the interstellar medium. VY CMa, in particular, is ejecting large amounts
of gas and dust at a prodigious rate. With its extended and complex
circumstellar nebula, it offers a unique opportunity to study its high mass
loss episodes and the detailed kinematics of this process.  

VY CMa is a special case even among the rare  hypergiants. It is
associated with a region of recent star formation, the young cluster
NGC 2362 and H II region Sharpless 310, and the dark cloud L1667, 
and has been considered a pre--main
sequence star (Herbig 1970a,b; Lada \& Reid 1978) as well as an evolved
massive star \citep{Hy69,RMH75,GW77}. 
The lack of Li I $\lambda$6707 \citep{Wall01} plus its powerful double--peaked maser 
emission, like other evolved stars, is strong evidence that it is a 
massive post--main sequence object. The measured $^{12}$C/ $^{13}$C ratio \citep{Ner89} however
is intermediate between what is measured for normal M supergiants and what is expected 
for the ISM. At a distance of 1.5 kpc \citep{GH72,LR78, 
Mar97} it is one of the most luminous red supergiants known 
($L \approx 5 \times 10^{5} L_{\odot}$). With its very visible asymmetric nebula 10$\arcsec$ 
across and high mass loss rate  ($2 - 3 \times  10^{-4} M_{\odot} yr^{-1}$) \citep{Dan94}, VY CMa 
is also one of the most important for understanding the high mass loss episodes in 
massive star evolution.    

Interferometric maps of the OH and H$_{2}$O emission reveal a complex structure interpreted
as either an expanding disk or bipolar outflow (Bowers, Johnston, \& Spencer 1983; Richards, Yates \& Cohen 1998) with expansion velocities of 30 -- 40 km s$^{-1}$ . Speckle imaging \citep{Witt98} revealed 
an asymmetric resolved component 0$\farcs$1 across,  
elongated approximately north-south. A geometry consistent with the maser maps and 
the  infrared images \citep{Monn99, Smi01} is an axially symmetric model with a 
slightly  flattened 
optically thick envelope or an equatorial disk and a bipolar flow/rotational  
axis aligned roughly northeast-southwest and inclined 15 to 30$\arcdeg$ to our line of sight 
\citep{MB80}. 

However, our multi-wavelength HST/WFPC2 images of VY CMa  revealed a complex
asymmetric distribution of material with no clear axisymmetric structures (See Figures 1,2, 
\& 3 in Smith et al. (2001), and with no evident  disk-like structure in the reflected 
optical light extending out to 2$\arcsec$ or more from the star as proposed by Herbig (1970a,b),
Efstathious \& Rowan-Robinson (1990), and Monnier et al (1999).
Instead, the circumstellar nebula is dominated by a prominent, nebulous  arc to the northwest 
 and two bright
filamentary loops or arcs to the southwest of the star, plus some relatively bright condensations 
or clumps of dusty knots near the star, all of which are evidence for multiple
and asymmetric mass loss events. The apparent random orientations of the arcs
suggest that they were produced by localized events not strongly aligned to
the star's axis or equator. The models for disks are based on modelling the spectral energy
distribution and therefore may be due reflected light from the prominent tail and the bright
condensations near the star rather than a true disk. The arcs and loops  also do not have 
the appearance of continuous
outflow streams as suggested by Monnier et al. (1999),  and are too massive
to  have been ejected by radiation pressure on dust. The initial ejection 
must have been induced by other processes.
We therefore  speculated that the arcs may be expanding loops 
caused by localized activity on the star's ill-defined surface, involving
magnetic fields and convection. Although this suggestion was  conjectural, large starspots do occur
on the surface of $\alpha$  Ori (Gilliland \& Dupree 1996; Kluckers et al. 1997; Uitenbroek, Dupree \& Gilliland 1998; Dupree, Lobel \& Gilliland 1999)  with accompanying 
outflows and chromospheric activity \citep{LD00,LD01}. Non-radial pulsational
instability is potentially an alternative ejection mechanism, but the distinction
between pulsation and convection  may be blurred in VY CMa.

To learn more about the morphology, kinematics, and possible origin  of VY CMa's 
complex ejecta, we have obtained long-slit spectra to map 
the emission and  absorption lines across several structures, including 
the prominent NW  arc and the two bright filamentary loops  to the southwest.   
Our results reveal the presence of a complex pattern of velocities in the
the strong K I emission line   and a significant velocity gradient 
across the dusty northwest arc. We find that;
\\  % line break for bullet mark
$\bullet$ Most of the gas and dust in the nebula, however,  is nearly stationary or moving 
very slowly with respect to the star. This is not only surprising but may be difficult
to explain for an evolved, massive  star. 
\\  % line break
$\bullet$ Embedded in this nebula are several discrete structures that are kinematically
distinct from the surrounding stationary nebulosity. The NW arc and the two prominence--like
arcs are expanding at 40 -- 50 km s$^{-1}$ relative to the star.
\\  % line break
$\bullet$ They were ejected in separate events over several hundred years from different
locations on the star.
\\  % line break
$\bullet$ A strong absorption line in the reflected spectrum in the ejecta is identified
with Ca I]. Although absorption from a semi-forbidden line seems unlikely, we demonstrate that
the identification is probably correct.\\ 
The strong K I emission in the star and the ejecta has never been adequately explained.
In the Appendix, we discuss a possible mechanism for forming the K I emission and 
other rare emission lines, including Ca I].

In the next section we briefly describe the observations and the data reduction 
procedures.  The spectrum of the central star and the origin of its  peculiar emission lines 
are discussed in \S 3 and in the accompanying Appendix. 
The kinematics of the gas and dust, the expansion of the arcs and loops, and the evidence
for stationary ejecta  are described in \S 4.  In the final two sections we summarize 
the evidence for localized high mass loss events and comment on the perplexing nature of VY CMa. 

\section{Long--Slit Spectroscopy and Data Reduction}

Spectra were obtained at four slit positions across the nebula (Figures 1a and 1b)
with the HIRES echelle spectrograph on the Keck1 telescope on Mauna Kea in December, 2002. 
The resolution was 40000 with a dispersion of .06{\AA} per pixel and a spatial scale of      
0$\farcs$19 per pixel. Our spectral coverage  from $\lambda\lambda$6410 to
8727{\AA} in 15 orders with inter--order gaps was chosen to measure the velocities of known 
 emission lines plus several absorption lines in the 8400 -- 8700{\AA} region. 
We used the B3 decker which gives a 14$\arcsec$ long slit with a 0$\farcs$57 slit width. 
Three of the slit positions are parallel across the nebula and  cross  
clumps of knots, filaments, and the NW arc. One of the slit positions(I) includes
the star. The fourth slit (V) 
crosses the two prominence--like arcs, 1 and 2,  in the outer parts of the 
ejecta to the southwest and the clump of knots south of the star.  
The seeing was estimated as  0$\farcs$8 or better throughout the night.
The observations are summarized in Table 1.

All of the data reductions and extractions were performed using IRAF\footnote{IRAF 
is written and supported by the IRAF programming group at the National 
Optical Astronomy Observatories (NOAO) in Tucson, Arizona. NOAO is operated by 
the  Association of Universities for Research in Astronomy (AURA), Inc. under 
cooperative agreement with the National Science Foundation}. The frames were  
trimmed, bias subtracted and flatfielded.
Calibration and object exposures were combined for each slit position. Using
the quartz exposures for reference, we extracted each echelle order from the 
combined frames using the `strip'  option in the IRAF task {\it apsum}.  
The dispersion corrections were determined for each order separately. No sky subtraction 
was done because emission and absorption lines are still apparent in the most distant 
apertures, beyond the visible nebulosity.   

The spectra were then extracted in 1{\arcsec} long apertures deliberately placed across
prominent features including the NW arc, the filamentary arcs, the knots and the embedded
star. Additional extractions were then made along the slit keeping 0$\farcs$4 separation 
between apertures. 
The positions of the apertures are summarized in Table 2; the positions are given  along each slit 
with respect to the point nearest the star, the radial distance of each aperture from the star, 
and the position angle from the star. Figures 1a and 1b show the positions of the extraction 
apertures for slits I and II, and for slits III and V, respectively. Four extractions
were made along slit III, across specific features listed in Table 2 plus four additional
apertures to the end of the slit. The wavelength calibration 
was then applied to each extracted spectrum. The separate exposures were  averaged to remove
cosmic rays, except for
those along Slit I which were summed because only two exposures were available.

\section{The Spectrum of the Embedded Star and Its Ejecta}

VY CMa probably represents the most extreme known case of stellar 
activity, in the same sense as, e.g.,  $\alpha$ Ori. 
Its inner wind (chromosphere?) has a density comparable to that
of $\alpha$ Ori, but the volume, mass, and energy are larger by
almost two orders of magnitude.  The photosphere is appreciably 
cooler than Betelgeuse and is about one order of magnitude more 
luminous.  The observed emission lines -- particularly the  strong  
K~I and the Ca~I] emission features -- differ from $\alpha$ Ori and have not 
been explained.  In the next few paragraphs we  review what 
one {\it expects\/} to see in this situation.

The photospheric temperature, luminosity, and radius are
thought to be, respectively, around 3000 K characteristic of an M5 type supergiant,  $5 \times 10^5 \; L_{\odot}$, 
and $R_{ph} \; \approx \; 2600 \; R_{\odot}$ $\approx$ 12 AU.  
(If  the wind is opaque, then $R_{ph}$ is larger than the 
true stellar radius $R_*$, the sonic point in the flow.) 
The usually quoted mass-loss rate and wind speed,
$4 \times 10^{-4}$ $M_{\odot}$ yr$^{-1}$ and 35 km s$^{-1}$, 
imply that the wind density at $r = 30$ AU is 
$n(H) + 4n(He) \; \approx \; 1.3 \times 10^{9}$ cm$^{-3}$ 
or  ${\rho} \; \approx \; 2 \times 10^{-15}$ g cm$^{-3}$. 

Given those parameters, we expect dust grains to condense near radius
$r_d \; \approx \; 100$ AU, at temperatures around 1000 K.  If the 
resulting dust-to-gas ratio is ``normal'', then, at red wavelengths
${\lambda}_{R} \sim 7000$ {\AA}, the opacity is 
${\kappa}_{R} \ \sim \ 170$ cm$^2$ g$^{-1}$ and the average 
optical thickness due to dust outside $r_d$ is
${\tau}_{R} \; \sim \; 50$.  Observations of scattered light
at radii $>$ 2000 AU seem consistent with this amount of dust \citep{Smi01}.  
With such a large average ${\tau}_{R}$,
one might naively expect the object to be seen only as a mid-IR 
source.  In reality, however, the dust should be very inhomogeneous.  
Even if the inner wind is smoothly isotropic (which we doubt), 
radiative acceleration acting on the dust near $r_d$ is unstable 
in a Rayleigh-Taylor-like manner;  the ensuing flow must be clumpy 
with a fairly small size scale.\footnote{
       %%%  footnote, re. size scale of dusty clumps: 
   Let $U$ = wind speed, $w$ = sound speed in the wind, and 
   $g_{rad}$ = acceleration due to radiation acting on the dust;
   then the three most obvious characteristic size scales for
   instabilities near $r_d$ are $w^2/g_{rad}$,    
   $({\kappa}_{R} {\rho})^{-1}$, and $r_d w/ U $.  For VY CMa,
   the last of these is of the order of 5--10 AU while the others 
   are much less;  so the expected condensations in the region 
   just outside $r_d$ should be, typically, no larger than 
   10 AU, and very likely smaller.}
       %%%  end of footnote
Therefore we should not be surprised to see visual- or red-wavelength
light from the star, escaping along low-density paths through a
very inhomogeneous dusty shell.  Since the material is expanding,
and a typical photon may be scattered one or more times before it 
escapes, the emergent spectrum is likely to be red-shifted by an amount
comparable to the wind speed $U$, and somewhat blurred in wavelength.  
Van Blerkom and Van Blerkom (1978) described this process for a constant 
wind speed, and the effect may be supplemented by radiative 
acceleration of material in the regions most favorable for photon 
escape, i.e., in relatively low-density gas between condensations.  

After escaping from the inner dusty region, some photons can be 
scattered sideways by dust in the outer ejecta, causing us to see 
an extended,  inhomogeneous halo around the bright central region.  
Any photon that is scattered toward an appreciably non-radial direction 
is Doppler shifted by a net amount $(1 - \sin \, \theta) \, U$, 
where $U$ is the outward wind speed and $\theta$ is  ($90\arcdeg - $scattering
angle).  Thus we expect the spectrum reflected in the halo to 
be more red-shifted than that seen when we observe the central 
source.  Since a line of sight through the halo typically samples 
a range of $\theta$-values, the reflected spectrum should be
blurred more than the central spectrum.

Generally speaking, observations of VY CMa confirm the above expectations. 
At wavelengths around 7000 {\AA}, very roughly 2 percent of the star's 
light escapes through the dust (see measurements reported in Smith et
al.\ 2001), suggesting typical optical depths of the order of 4 or 5, 
not 50, along the photon escape routes.   Absorption lines in the central 
emergent spectrum are Doppler-shifted by about +25 km s$^{-1}$ relative
to the supposed systemic velocity, and are somewhat broader than one 
expects for such a low-gravity star.  When reflected in the halo, the 
same features have larger Doppler shifts and are more blurred (Fig. 6).  
Throughout this paper we shall note additional details that do {\it not\/} 
fit the above picture in straightforward ways.  

\subsection{The Spectrum of the Central Star}

The K I resonance lines ($\lambda\lambda$ 7665, 7699) are not only the strongest emission lines 
in the spectrum of VY CMa \citep{Wall58}, but  these lines, rarely observed in emission, 
are  stronger in VY CMa than in any other known late-type star (see Bernat \& Lambert 1976, Guilain \& Mauron 1996).  In addition to K I,  strong emission 
is also  present in the   Na I  ``D'' lines,  Ca I] $\lambda$6572{\AA}, and Rb I 
($\lambda\lambda$7800, 7948)\citep{Wall71}.  Narrow
band heads of ScO (Wallerstein in Hyland et al 1969 and Herbig 1974), and  
band heads of TiO and VO, plus lines of Fe I, Ti I and Cr I \citep{Wall86} are also in emission.
Our wavelength coverage includes  K I $\lambda$7699,                                     
Ca I] $\lambda$6572 and Rb I $\lambda$7948, but not the Na I ``D'' lines.
In addition to the emission features, there are
strong absorption lines of Ca II, Fe I, and Ti I in the wavelength region from
approximately 8400 to 8700{\AA} that are relatively free from molecular
absorption bands so they  can be confidently used for radial velocity measurements.
In our data, the central star is in aperture 2 on Slit I (Fig. 1a), and  examples of the 
emission and absorption lines in the ejecta are shown in figures throughout the paper. The 
spectrum at the star does not show any significant differences from those published earlier 
and the velocity 
measurements for the star summarized in Table 3 are consistent with previously published results.  
All velocities in this paper are Heliocentric.

In Figure 2 we show the K I and Na I line profiles on the star from a HIRES spectrum 
obtained by George Herbig in 2000 that includes the D lines. Both the K I and Na I emission have peak 
fluxes several times as high as  the continuum, and not surprisingly,  their doublet ratios show that 
they are optically thick. The minima in the strong P Cygni profiles have the same velocity
at $\approx$ 15 km s$^{-1}$. For the K I lines, this is a shift of 
$\approx$ 30 km s$^{-1}$ from the emission peaks which have  a mean velocity of 45 km s$^{-1}$, 
consistent with our measured K I and Ca I] emission velocities. The Na I emission peaks 
however  occur 
at $\sim$ 60 km s$^{-1}$ which  agrees with our mean absorption line velocity
at the star (Table 3).  It has been recognized for some time \citep{RMH75, GW77} 
that the absorption line  velocities  are not only different from the 
K I and Ca I] emission lines, but also deviate
significantly from the velocity of the system inferred from the OH and H$_{2}$O maser observations, 
V$_{LSR} =$ 20 -- 22 km s$^{-1}$ (V$_{Hel}$ $\sim$ 35 -- 37 km s$^{-1}$)(Bowers, Johnston \& Spencer 1983).  Herbig (1970a) first suggested that the stellar absorption lines are red-shifted and broadened due
to scattering by an expanding dust shell or equatorial disk, and as mentioned above, this was 
modelled by Kwok (1976) and van Blerkom and van Blerkom (1978). Note that our velocities 
for the K I and Ca I] emission lines at $\sim$ 41 km s$^{-1}$ are significantly less than 
the mean absorption line velocity. This difference is usually attributed 
to the location where  the emission lines are formed, outside the inner radius of the 
dusty shell or disk.    

The extremely strong K I emission and  other rare emission lines 
such as  semi--forbidden Ca I] are difficult to explain. 
In the Appendix we discuss problems associated with the formation of the K I lines 
in VY CMa. Some  obvious excitation processes
such as collisional and radiative recombination are eliminated, and while  resonant scattering  
is the most promising explanation, it requires a special geometry to cause  
strong net emission. We propose a model for the 
production of the very strong K I emission observed at the star, but problems 
associated with resonant scattering remain, especially in the ejecta.

\subsection{The Reflected and Scattered  Spectrum in the Ejecta}

The strong K I emission line is our best tracer of the gas in VY CMa. It shows 
considerable variability with multiple emission peaks 
throughout the ejecta (see also Smith 2004). Examples of the line profiles  
along Slits I, II and III are 
shown in Figure 3, with their velocities  summarized in Table 4.\footnote{The emission 
line profiles along Slit V are in  Figure 7 and their velocities in Table 6.} The emission 
profile on the star  is single (Slit I, Ap 2) with a strong P Cygni
absorption feature formed in its expanding wind. The P Cygni profile is present in the 
reflected spectrum near  the star and is gradually red-shifted due to scattering by the
dust expanding outwards.  Off the star, the K I profiles 
along  the slits are much more complex, often with two or more recognizable emission peaks.  
In the outermost apertures  the profiles 
become single and narrow. An emission peak with the expected systemic velocity is observed
throughout  much of the ejecta to the west and northwest of the star. It has a nearly  
constant velocity at increasing distances
and is essentially motionless relative to the star. This and other
evidence for stationary gas and dust in VY CMa's ejecta are discussed in \S 4.3.  

Interestingly, a broad and shallow K I absorption feature to the red of the emission 
profile is present at positions where the slits cross the dusty condensations and 
nebulous arcs primarily to the west and NW of the star.  
This may be an inverse P Cygni profile, but the broad absorption minimum has a 
red-shifted velocity, typically $\sim$ 86 km s$^{-1}$ (Table 5)  like the other 
reflected absorption lines at the same positions (see \S 4.1). Note that on average, the
 K I emission greatly exceeds the strength of the absorption in our data. One would 
normally expect emission to balance absorption for a resonance--scattered line (see the Appendix).

The semi-forbidden Ca I] $\lambda$6572 emission line (Figs. 4 and 5) is visible only in the 
extractions on and  near the star. The line is very narrow, does not show the 
multiple peaks observed in the K I line,  and has essentially a constant Doppler velocity at  
$\approx$ 41 km s$^{-1}$ like that for the
K I emission on the star (Table 4).   A strong, reflected  absorption line which we attribute 
to Ca I] absorption is present in the ejecta. Although absorption in a semi-forbidden line
seems unlikely, we demonstrate later, \S 4.1, that this identification is almost certainly correct.  
Another rare emission line, Rb I at $\lambda$ 7948, is present in the reflected spectra,
but is relatively weak and broad.

H$\alpha$ emission in VY CMa is usually reported as very weak and variable. We find that the relative
strength of the H$\alpha$ emission depends on position in the nebula (See Figs. 4 and 5). In the inner
region of the ejecta it  is weak and broad with an associated  broad shallow absorption. 
The narrow H$\alpha$ emission line visible in the outermost extractions along all of the slit 
positions is nebular from the nearby H II region Sharpless 310. VY CMa lies just to the west of an
arc of nebular emission. Weak nebular emission from [N II] and [S II] 
is also present in the spectra at the same positions.  The  velocity of this  
H$\alpha$ line at $\approx$ 35 -- 37 km s$^{-1}$ agrees with the systemic velocity and  the 
velocities of the  K I lines  measured at the same positions, thus
confirming the close association of VY CMa with the cluster and H II region.  

\section{Kinematics of the Gas and Dust} 

The reflected and scattered spectrum in the circumstellar ejecta reveals a remarkably complex 
behavior of both the emission and absorption lines. The K I emission
line profiles and their Doppler velocities vary with position in the nebula 
as the slits cross recognizable knots, arcs, and filamentary features in the image. 
They also indicate apparently stationary gas in the same projected locations,  which is
difficult to understand for the reasons noted below.
The reflected absorption lines  show a strong velocity gradient across the 
prominent nebulous arc to the NW of the star, while the same absorption lines along
slit V show  virtually no
variation in Doppler velocity. These  phenomena are discussed  
in the following subsections.

\subsection{The Velocity Gradient Across the NW Arc}

The absorption lines from the obscured star and the dusty ejecta are  reflected or  
scattered by the embedded dust grains and reveal both the relative
motions of the dusty clouds and the velocity dispersions within them. The mean 
velocities of the  strong absorption lines of Fe I, Ti I, and Ca I
in the far--red spectrum are  summarized in Table 5 
and examples of their profiles are shown in Figure 6.

A few of the lines in apertures on and near the NW arc are double or slightly asymmetric. 
For example, four of the absorption lines in I Ap 3 are double 
so we used the blue components, which agree with the velocities of the single lines, 
for the mean velocity in Table 5. In the extractions immediately on either side of the 
NW arc, a few of the absorption lines also have asymmetric profiles. Only the single lines  and 
well defined absorption minima of the some of the asymmetric lines were used for the mean
velocity.  The asymmetries may be  due to additional dusty clouds along the line of sight
or to the expected gradient along the NW arc.

The absorption line profiles from the dusty arcs are  broad and shallow compared to the 
narrower profiles  on and near the star.  Along Slit I, for example, the equivalent widths
of the absorption lines show only a small decrease from the star to Ap 3 and Ap 4 (NW Arc), 
but the FWHM of the same profiles increase by a factor of
2 to 3 due to the scattering and motions of the grains. To determine the velocity 
dispersion of the particles responsible for the broadening, we smeared the stellar 
profiles of the Fe I and Ti I absorption
lines with a Gaussian. We used a range of values of $\sigma$ from 0.5{\AA} to 1.6{\AA}
and compared the smeared stellar lines with the observed profiles along the slits.
The best fit varied slightly from line to line and of course for the different 
apertures, but for those profiles across the NW arc, the required smearing corresponds 
to a FWHM of 85 to 120  km s$^{-1}$.  The total velocity range will be even larger. The dusty scattering regions in this visible feature
thus have a relatively high velocity dispersion in addition to their net motion with respect
to the star.

The  mean velocities  for the absorption lines are plotted in Figure 7 with respect to the 
distance along each slit from the point nearest the star.  They show a large redshift  as 
the three parallel slits (I, II and III) cross the
dusty arcs to the west and NW of the star,  and then decrease in the outer nebulosity
beyond the prominent NW arc.  {\it The profiles across the NW arc are significantly 
red-shifted with a velocity difference of $\sim$ +50 km s$^{-1}$ relative to
the star, suggesting a net motion of the expanding arc of this order.} 
This difference is presumably the ``moving mirror'' redshift, $(1 - \sin \, \theta) \, U$,  
where U is the outward expansion of the reflecting material relative to the star 
and $\theta$ represents the direction of the flow
relative to the plane of the sky ($\theta > 0$ for motion toward us). 

If the NW arc represents the material from a single ejection event we expect the 
expansion velocity to be proportional to distance  along the arc. This 
expected trend is not immediately obvious in the observed velocities. 
It is probably blurred in our data because of
the size of the extraction apertures, although  
some subtle evidence for this effect may be present
in the velocities between Slits II and III in apertures directly on the arc; 
III Ap 4 is slightly further from the star and has a  higher velocity than II Ap 5.  
Moreover, in Figure 6, one can see a velocity component on the short wavelength
side of the lines at position II Ap 4 and on the long wavelength side at II Ap 6.
The velocity difference is consistent with the uniform expansion expected for the NW arc.

Unlike most other locations in the ejecta, 
appreciable K I {\it absorption} is observed in the extractions on and near the NW arc.  
It also shows an appreciable redshift 
at  $\approx$ 85 km s$^{-1}$ (Table 5), and the line is 
also quite broad and asymmetric. Unlike the other absorption lines, the K I velocity is
approximately constant across the arc. This is probably due to the strong emission line 
covering or 
overlapping part of the absorption feature. The red wing of the K I absorption, however,  
does show a redshift as the slits cross the arc, from 135 to 151 km s$^{-1}$ along Slit II.

Using the velocities from the absorption lines at position II Ap 5, 
directly on the NW arc,  we apply its ``moving mirror" redshift of 54 km s$^{-1}$ to 
the K I velocity at this position, and get 
31 km s$^{-1}$, close to the   35 km s$^{-1}$ systemic velocity of the star. 
 We therefore assume that the K I absorption in some sense represents the 
spectrum of the star as seen from the NW arc. In Figure 8 we show the K I and Ca I] 
line profiles shifted by (35 + 54) km s$^{-1}$,  one of the ordinary absorption lines,
and the K I emission at the star's position, I Ap 2. The  K I and Ca I] absorption
features are centered close to zero and are thus nearly in the rest frame of the
star. The Ti I absorption line is shifted an additional 25 km s$^{-1}$ in Figure 8, because as
explained in \S 1, it arises inside the circumstellar dust formation region.  
Since the K I and Ca I] absorption features do not need the extra 25 km s$^{-1}$ shift, 
we conclude that they occur {\it more than 100 AU from the star,
outside the ring or shell of dust.}\footnote{Since Ca~I] $\lambda$6573 is a 
semi-forbidden line, one 
   would not have expected it to appear in absorption as
   observed.   But the wavelength coincidence is essentially
   perfect, and we have found no other likely identification 
   for the observed feature.  Thus we must conjecture that some
   atomic-physics trick makes the oscillator strength much
   larger than one would expect. } 
The red wing of the K I
absorption, at $\approx$ 60 km s$^{-1}$ with respect to the absorption minimum,  indicates the
presence of a  substantial dispersion within the
expanding gas of 100 km s$^{-1}$ or more, about the same as observed for the other 
absorption lines at the same locations. This means that either the NW arc has a large
disperison in its expansion velocity or a large extent along the line of sight, or both.
 Note that
the ``terminal velocity'' of the K I P Cyg profile as seen at the star is also
about 45 -- 50 km s$^{-1}$.
Thus, the moving-mirror redshift at the NW arc, the P Cyg ``terminal velocity'', the red wing of the
K I absorption,  and the reflected line
widths {\it all suggest flow speeds of about 50 km s$^{-1}$}. Expansion velocities greater than
35 km s$^{-1}$ are required by the absorption line widths and there is no indication of velocities 
much greater than 70 km s$^{-1}$ relative to the star.

An expansion speed of 50 km s$^{-1}$ implies that $\theta$ must be near zero so the NW arc 
is basically moving across our line of sight. This is not a surprising result given its
orientation and appearance within the nebula. Furthermore, the limits on the expected expansion
velocity discussed above, 35 to 70 km s$^{-1}$, would restrict $\theta$ to between
-25$\arcdeg$  and +17$\arcdeg$. Assuming mostly transverse motion for the arc, at its distance
from the star of about 3$\arcsec$ or 4500 AU at 1.5 kpc from the Sun, the material in the  
{\it NW arc was ejected about 400 years ago} (-100, +200 yrs.). 

We get a somewhat different result for slit I which crosses the NW arc at a
section where it appears to be bending back toward the star. At I Ap 4 the Doppler 
velocities of the absorption lines  are slightly lower and the ``moving mirror" redshift 
relative to the star is 40 km s$^{-1}$.  Assuming  50 km s$^{-1}$ for $U$, 
$\theta$ is roughly  $+11.5\arcdeg$ for
this position. Thus, this section of the NW arc may actually be the nearer side  
with a small component of motion toward us.  

The  corresponding K I emission, positions II Ap 5, II Ap 6, III Ap 4 and III Ap 5, has 
two definite velocity peaks (Figure 3) which can also be seen as two separate emission 
features in the image of Slit III in Figure 9. The shorter wavelength one presents some problems that are
discussed in the next section. The red-shifted component with  velocities of  60 to 70 km s$^{-1}$ exhibits a small velocity gradient across the arc as one would expect for an ejection. 
Recognizing this trend, we
can identify a corresponding bump in the line profiles at II Ap 4 and III Ap 3 
at $\sim$ 52 km s$^{-1}$ that probably corresponds to the same gas.  
Since this feature spatially coincides with the NW arc and is expanding, we presume that 
this is  the associated gas. However, the Doppler
velocity, at II Ap 5, for example, corresponds to a shift of 25 km s$^{-1}$ relative to the star,
half that for the absorption lines. The slope of the velocity gradient, 5 km s$^{-1}$ per 1000 AU 
yields an expansion age of 950 years. If this emission is due to resonant scattering, then 
it is moving away
from us at an angle of $\sim$ 30$\arcdeg$ which sems unlikely given the results from the absorption
lines in the NW arc. If, on the other hand, it is   reflected emission,  it will be  
directed toward us  at $\sim$ 30$\arcdeg$ since the redshift is smaller than for the 
absorption lines. The expansion age for the emitting gas is then about 550 years, 
comparable to the above results for the NW arc. 

Slit positions III Ap 3 and II Ap 4 cross a filamentary, somewhat twisted arc just west 
of the star, between it and the NW arc. If we apply the same procedure described
above for the NW arc with the K I and Ca I] absorption lines, we find that the moving
mirror velocity for the dust is 45 km s$^{-1}$ and the flow speed is 55 km s$^{-1}$ with
$\theta$ of $+10.5\arcdeg$. At its distance from the star, this material was ejected only 200
years ago.

In contrast with the results for the ejecta to the west and NW of the star, 
the Doppler velocities of the reflected absorption lines along Slit V are nearly 
constant (Table 5), show no gradient with position (Figure 7), and no significant 
velocity shift with respect to the velocity at the star.  This suggests that the reflecting material 
is basically stationary relative to the star. This and other evidence for surprisingly stationary 
gas and dust in the circumstellar nebula is discussed in \S 4.3.  

In the next section we discuss the  significant velocity shifts in the scattered light 
from some of the K I emission lines in the ejecta and additional evidence for ejection 
of discrete features in the nebula.

\subsection{Multiple Velocities in the K I Emission Feature}

The K I emission line often shows a complex profile with multiple peaks
that vary with position in the ejecta. If  the K I emission lines in the ejecta are
due to resonant scattering, the Doppler shifted and multiple peaks observed
at different places  must be due to gas moving at these different
velocities thus permitting us to determine the expansion of the nebula and motions within it. 
However reflection by the dust can also contribute to the observed profiles. In this case we 
would expect to see either a  single profile like that on the star, or an emission peak, with  a 
velocity shift between the  emission velocity at the star and the absorption lines at the 
same position. 
A few of the K I profiles closest to  
the star may  be due to reflection or have a contribution from reflection. These 
appear to be either single or lack obvious multiple peaks even though the lines are 
quite broad. Some possible examples 
are the  K I emission at  I Ap 1,  V Aps. 1 and 2 and  III Ap 2 in Figures 3 and 11. 
For the multi-peaked profiles  further from the star, it is  difficult to determine  
if the emission is  due to resonant scattering or reflection. 
Both processes are probably occurring. 

In Figure 10 we show the variation of the emission line velocities along the different
slits as we did in Figure 7 for the absorption lines. Slit V is shown separately.
Along Slits I, II and III we see a pattern to the west and northwest of the star 
in which most of the K I profiles have
an emission peak at a velocity near 35 km s$^{-1}$ ( see Table 5).  
This is very apparent in Figure 9 as the strong, straight emission
feature in the two-dimensional image of Slit III.   
Whether these emission peaks and single profiles are due to reflection or to resonant 
scattering, their velocities are consistent with stationary gas,  near zero velocity 
with respect to the star. This possibility  is discussed more fully  in \S 4.3. 
A second, somewhat weaker emission peak  present in extractions on the NW arc 
was discussed in the preceding section.
 
The changing profile and multiple peaks of the K I emission line along Slit V (Figure 11) 
show the most dramatic shifts in Doppler velocity with position in the nebula, and 
if these are due to resonant scattering in the gas, imply the existence of clouds or 
filaments of gas with a wide range of velocities in the ejecta. 
The K I line varies from a single profile in Ap 1 with a strong 
P Cygni absorption, similar to that viewed on the star,  to multi-peaked profiles with large 
velocity shifts where the slit crosses knots and filaments in the outer parts of the nebula.  
The K I velocity to the north and east of the star,  V Ap 1 and
Ap 2, is consistent with that observed on the star and may simply be the reflected 
profile from the star as we mentioned earlier.  In apertures south of the star  across the 
dusty knots and arcs, one of the
strong emission peaks in apertures 3 to 6 has a velocity like that of the reflected absorption lines 
(60 -- 65 km s$^{-1}$). 
This  emission component may be reflected by the dust or due to resonant
scattering by gas moving  $\sim$ 30 km s$^{-1}$  relative to the star. Likewise, 
the blueward peak near 25 km s$^{-1}$ in the same apertures, may be either reflected or
scattered emission and represents a separate flow of emitting gas along the same line of sight. 
The most striking change occurs across the two outer
filamentary arcs (apertures 5 to 7) with the appearance of a strong  blue-shifted emission feature.
The  sharp transition from an  emission  peak near 20 km s$^{-1}$ in aperture 5 on arc 2 to 
the  blue-shifted emission
across arc 1 at $\approx$4 km s$^{-1}$, suggests that this is resonant scattering 
in kinematically and spatially separate gaseous filaments or gas flows moving very differently from the reflecting dust. 
This shift can be clearly seen in the   two-dimensional image in Figure 12  where 
narrow emission appears in the P Cyg absorption core and broadens out across  arc 1.
The  P Cyg absorption is either no longer present  or has been blocked by  the strong blue-shifted
emission. This is not a continuous outflow  from a single ejection as suggested by Smith (2004),
 because these two emission features are  resolved in the image of the slit.
The velocities of the emission peaks associated with these three features
are shown connected by a lines in Figure 10.

We can trace this  prominent blue-shifted emission feature in the two dimensional image  
over $\approx$ 3$\arcsec$ in the ejecta and directly across arc 1 in aperture 7.  It  
corresponds to the strong emission peaks in apertures 6 and 7 and may also be weakly 
present in Ap 8 and possibly in Ap 5 as a small bump on the blue side of the profile. 
In Ap 6  broad, weaker emission scattered or reflected in gas at a range of velocities up
to 60 km s$^{-1}$ is also present, but is not observed in Ap 7 across the arc.
We therefore identify this strong blue-shifted emission feature with resonant 
scattering by gas in a stream or flow associated with the material in arc 1.   
The  Doppler velocities  at  4 -- 8 km s$^{-1}$ indicate that this material is moving  
relative to the star at $\sim$ 30 km s$^{-1}$  toward the observer.
This of course contrasts sharply with the nearly constant 60 km s$^{-1}$ velocity of 
the absorption lines, and suggests a significant kinematic difference between the gas in 
the arc and the surrounding circumstellar material similar to what we found for the NW arc.
Although arc 1 has a significant velocity toward us relative to the star, its motion 
across our line of sight is not known. If we adopt 30 --  35 km s$^{-1}$ from the width
of the broadened absorption lines (see \S 4.3) at the same positions, as representative of the motions
in the surrounding material  for the transverse velocity,   
then the total motion relative to the star may be  as much as
45 km s$^{-1}$. The material associated with arc 1 would then have been ejected 
$\sim$ 1000 years ago assuming it is moving in the plane of the sky.  

The two prominent emission peaks at 22 and 65 km s$^{-1}$  across arc 2, in aperture 5, 
are well separated kinematically in the two dimensional image
of the slit, but a spatial separation or identification with distinct features is not 
obvious in the images at these positions. The 65 km s$^{-1}$ feature agrees with the velocity of the absorption lines in the 
same aperture, so we  assume that   this gas is associated with the reflecting dust,  and 
since it is more red-shifted that it comes from background material. 
The second emission peak  may then be due to reflection or resonant scattering by gas in arc 2. 
This implies a motion of $\approx$ 13 km s$^{-1}$
relative to the star for the emitting gas along the line of sight.
The corresponding motion of the arc across
our line of sight is not known and may be different from the assumed 30 -- 35 
km s$^{-1}$,  but with this motion, arc 2 is moving at $\sim$ 37 km s$^{-1}$ and was ejected about
800 years ago. Given the uncertainty in these estimates, it is possible that arcs 1 and 2
were ejected at about the same time. However the slit image shows that they are kinematically
separate emission features which are also spatially separate in the WFPC2 image. Therefore
assuming comparable outflow velocities relative to the star, arc 1 would have been ejected
earlier, but the ejection times  also depend on the projection angle for these 
structures which is not known. 

Other distinct features in the ejecta such as the two clumps of dusty knots, S and SW, are
much closer to the star and presumably represent material ejected much more recently. 
The evidence for localized mass loss events and stellar activity are discussed in
\S 5.

\subsection{Motionless Gas and Dust in the Ejecta?}

The presence of gas and dust throughout the nebula, apparently at the same velocity of the star, 
presents a serious problem because it implies that much of the circumstellar material is 
nearly stationary or moving very slowly relative to the star.
This certainly conflicts  with  what one expects for a red supergiant with a high 
mass loss rate.  
 
In the ejecta to the west and NW of the star, the K I emission line  has a velocity component 
consistent with relatively motionless gas, at the systemic velocity of the star, 
extending over $\approx$ 6$\arcsec$ or $\sim$ 9000 AU in the  nebula (Figure 10). 
If this emission is due to reflection by dust, we should  see a  shift
in velocity with respect to the rest frame of the star even if   the motion is 
transverse from a uniform flow of gas outwards.  This is not observed.  
If it is resonant scattered emission, and if the emission comes from widespread transparent
gas,  we would expect to see a broad profile centered at the average velocity of the system. 
However,  much of the nebula where this K I emission is observed is  not 
transparent and,  more importantly, the emission components are quite narrow. The FWHM's range from
0.25 to 0.4 {\AA} or only  10 -- 15 km s$^{-1}$ wide, compared to $\sim$ 1{\AA}
for the single K I profiles on and near the star. This is much less than the expected expansion
speed of the ejecta and  suggests that there is very little Doppler broadening in the lines.  
{\it The emitting gas is therefore either not moving relative to the star, or all of the material 
must be moving in the plane of the sky} i.e. in a narrow plane viewed perpendicularly.  The latter 
seems highly unlikely over such a large area.  Furthermore,  some of the profiles show two
emission peaks, one near 35 km s$^{-1}$ and the other at 41 km s$^{-1}$, the velocity at
the star, suggesting that the emission is due to both reflection and resonant scattering. 
The narrow K I emission lines in 
the outer ejecta also have very broad emission wings that extend to 100 -- 200 km s
$^{-1}$ from the center of the emission line (Figure 13).  They are formed too far 
from the star to be electron scattering wings. It is possible that these emission 
wings are from more distant gas and dust, expanding outwards from VY CMa perhaps in a 
more spherical distribution from a much earlier unstable period in the star's evolution.

There is also some evidence for constant velocities, or zero motion relative to the star, 
in the reflected light to the 
northwest of the star. After crossing the near side of the NW arc, the absorption 
line velocities measured in  the outermost apertures along Slit I have a velocity like 
that observed at the star (see Figure 7). Along Slits II and III, however,  the absorption
line velocities decrease, but stay red-shifted relative to the star by $\approx$ 20
km s$^{-1}$ perhaps because the material, in the same plane,  is being pushed by 
the expanding arc.  

The Ca I] emission has a nearly constant velocity at 40 -- 41 km s$^{-1}$ throughout the
ejecta. Since this is the same as its velocity on the star, we assume that this is 
reflected emission. The CaI] emission profile does not show the variability with position
and multiple peaks readily apparent in the K I emission line. As mentioned in the Appendix,
the presence of this semi-forbidden line in both emission and absorption is not understood. 
 
Along Slit V we observe absorption lines, including Ca I], and  some K I emission  
with nearly constant velocities 
extending over much of the length of the slit and across the nebula\footnote{Except for Ap 1, there is no K I absorption along Slit V.}. All of these
indicators have apparent Doppler velocities near 60 -- 65 km s$^{-1}$ which is also the 
red-shifted velocity of the absorption lines scattered by the dust envelope
around the star. For the absorption lines, this implies that the reflecting dust 
is not only barely moving  but also has little motion  with respect to the star. 
Unlike resonant scattered emission, reflected absorption lines are expected to show a redshift 
relative to the star even if the reflecting dust is moving perpendicular to our line of sight. 
Admittedly, the
problem is reduced if the material is moving toward us; in that case, the ``moving-mirror'' effect 
is less, but there should still be a measurable redshift. Two selection effects, forward
scattering by the dust plus extinction,  can also serve to reduce the measured redshft because
we  preferentially see the side of the ejecta toward us. But these considerations do
not explain the near constancy of the velocities with distance from the star and the lack
of any additional shift in the apertures across arcs 1 and 2, which are quite red and
dusty in the images. 

The absorption line profiles along Slit V are also broadened compared
to those on the star, although not as much as across the NW arc (see Figure 6). We compared the Gaussian smeared stellar
profile with the observed profiles and obtained a range of results which varied with
position along the slit.  The reflected profiles from apertures 2 and 3 
closest to the star, showed only marginal broadening with respect to the star, 
even though aperture 3 crosses the relatively bright dusty knot or condensation 
just south of the star. The smeared profiles that best matched
those observed from aperture 4, which crosses several small
arcs,  have FWHM's corresponding to 20 - 40 km s$^{-1}$, and for apertures 5 
through 7 across the outer filamentary arcs 1 and 2, the FWHM's are 65 km s$^{-1}$ .
Although this is smaller than measured for the NW arc, it represents  significant 
motions along the line of sight of $\pm$  30 -- 35 km s$^{-1}$ for the material within these
arcs, in contrast with the  apparent lack of motion with respect to the star.
Consequently,  we would expect to see a redshift relative to the star of at least 10  km s$^{-1}$ or
so, especially across the arcs even with the above considerations for a reduced redshift.
Instead, the motion relative to the star is at most $\sim$ 5 km s$^{-1}$ assuming that 
the material along Slit V sees the same reflected and red-shifted spectrum of the star that
we see, I Ap 2. 

It is  possible that at different positions in the  nebula, we may be observing radiation
that escaped from  different regions on the star, perhaps through holes or gaps in the dusty envelope.
Consequently, the gas and dust at these locations may be getting a different view or 
spectrum of the star.  For example, if there is a NE-SW axis of symmetry 
with a bipolar flow, as has been suggested, then along Slit V we may be observing  
gas and dust that views the star from near the polar direction which would be less obscured if 
 there is a disk or flattened dusty envelope around the star.   
If gas is escaping 
through a low density region in the dust envelope, then the reference velocity for 
the reflected absorption lines would more likely 
be closer to the systemic velocity than the  red-shifted velocity observed at the star. If this is 
the case, the reflected material may be moving at 30 km s$^{-1}$ relative to the star.
This is just a hypothesis, but it might solve the problem of the zero-velocity gas and dust 
along Slit V. However, it would also require that along most of Slit V, we see the star
from exactly the same direction or angle which then poses problems for the assumed geometry. 

A person unfamiliar with the nature of VY CMa would conclude from the above description
that a substantial amount of nearly stationary or very slow moving gas and dust exists 
in its ejecta. 
This is obviously a paradox given the star's strong wind and probable mass loss history.
At this time we have no satisfactory explanation. It is possible that the general flow of 
the diffuse gas in the extended circumstellar material  is extremely slow, but the escape velocity 
for VY CMa is $\sim$ 60 km s$^{-1}$ at its photosphere 
and 20 km s$^{-1}$ at the dust formation radius at 100 A.U. Conceivably this material has been 
slowed by some process or it did not have quite  
the escape velocity and is now falling back. But there is no  evident reason
for this to happen, other than it would help solve our paradox.

\section{The Evidence for Localized Mass Loss Events }

The results of our high resolution spectroscopy demonstrate that the 
three primary arcs in the visual images of VY CMa  are kinematically distinct 
from the surrounding stationary circumstellar material. The NW arc and arcs 1 and 2 
have expansion speeds of 50  km s$^{-1}$ to $\sim$ 40  km s$^{-1}$, respectively. 
The gas and dust associated with these
structures was thus ejected in specific mass loss episodes; $\sim$ 400 years ago for the 
NW arc and possibly 1000 to 800 years ago for arcs 1 and 2. For the ``western arc'' we found an 
expansion age of only slightly more than 200 years. Thus we have evidence for a period
of considerable activity during the past 1000 years, including some recent ejections.
In addition to the arcs, numerous clumps of small knots or condensations are observed in the 
images relatively close to the star. The dusty clumps to the south(S) and southwest(SW)
of the star are the brightest, but several very small knots are observed just to the
west and northwest of the star. These are clustered quite close, within 0$\farcs$5 of
the star, and must have been ejected quite recently, only $\sim$ 70 years years ago assuming 
an outflow speed of 50  km s$^{-1}$.  

With the possibility of ejection episodes this recently, there may be 
some hint of activity in the  star's observational record. Numerous ground-based observers in
the past have often described the appearance of knots of nebulosity that come and go and
may vary in position, but the timescales for the motions and variablility seem too short
to be related to the expanding ejecta. (See Smith et al 2001 for a discussion and the references)
These variations could be due to changing illumination
of the dusty ejecta by radiation escaping from gaps or low density regions in the envelope
\citep{Wall78, Smi01}. 
VY CMa's historical light curve \citep{Rob71} shows some very intriguing variability. 
The most significant  change is its decline from
m$_{v}$ $\sim$ 6.5 mag. beginning in 1872 to $\sim$ 8 mag. by 1880 with some brief brightening 
episodes superposed. Whether this  
1.5 mag decrease was due to a change in bolometric correction or to the formation of dust, 
is not known, but it may correspond to the appearance of several knots and small arcs about 100 to
200 years ago. The star apparently then slowly faded to about 8.5 -- 9th magnitude, and 
although the record is spotty, it may have had periods of greater variability in the 1910's 
and again in the 1930's. The more modern record based on the AAVSO light curve and published 
photometry, shows
that VY CMa displayed considerable variability of 1 to 2 magnitudes from about 1984 to 1995 with 
oscillations lasting several 100 to a thousand days that resemble the pulsations in 
irregular variables. Since then it has been relatively steady at about 8.5 mag.  

The NW arc is the most clearly defined structure or mass concentration within the ejecta
and is therefore the best candidate for a mass estimate. In our imaging paper, we estimated its
total mass at $\sim 3\times 10^{-3}$ M$_{\odot}$ from the surface brightness in a well resolved 
2 arcsec$^{2}$ section. In this estimate we assumed an optical depth of unity. This assumption
plus the evidence from the velocity gradient across the arc that more material may be 
participating in the outflow than is apparent in the images, suggests that this may be an
underestimate for the arc's total mass. The dynamical timescale for VY CMa is on the order of three 
years as are the episodes of large  variability in its light curve. So whether the observed 
variability and the ejection events are due to non--radial pulsations or large 
scale convection and activity on the star, the timescales will be comparable. Therefore the 
short term mass loss rate associated with the ejection of the NW arc was at least 10$^{-3}$ 
M$_{\odot}$ per year. The corresponding kinetic energy of the NW arc is $\sim$ 10$^{44}$ ergs
which is not unrealistic for a star like VY CMa with a total energy output of 6$\times 10^{46}$ 
ergs per year.  

Given the evidence we have for a relatively stationary nebula, most of VY CMa's mass loss may be
occurring in these separate high mass loss events perhaps due to activity on its surface. If we 
associate the ejections and surface activity with VY CMa's photometric
variability then they may occur approximately every 50 years or so. The NW arc may be an especially 
massive outflow, but if we assume $\approx$ 10$^{-3}$ M$_{\odot}$ per ejection, then over a thousand
years the mass loss rate would average $\approx 2\times 10^{-4}$ M$_{\odot}$ per year 
which is close to what we estimated in Smith et al (2001) for the entire circumstellar nebula and
to previous independent mass loss measurements \citep{Dan94}. 

The random orientations of the three primary arcs relative to the star together with
their different ages, suggest that they were ejected not only at different times, but by
localized events or disturbances at different positions on the star.
Our results show that the NW arc is moving primarily across our line of sight with
little projection with respect to the plane of the sky.  VY CMa's geometry is unknown,  
but if we assume the NE-SW rotation/symmetry axis from the maser model, then the NW arc
is very likely within 10$\arcdeg$ to 15$\arcdeg$ of the expected equatorial plane.
Given the relative positions
of arcs  1 and 2 within  the nebula and with respect to the presumed axis,  each has likely
been ejected from different locations on the star, at higher latitudes relative to the
NW arc, and possibly from the polar region, but that depends on the unknown projection angle.

One of the  uncertainties in the above  discussion is our lack of information on the transverse
motions of the ejecta and the geometry of the nebula.  We will soon be obtaining second epoch
images and imaging polarimetry of VY CMa with HST to measure its expansion and map the 
morphology and relative distribution of the embedded structures.

\section{Concluding Remarks On the Nature of VY CMa} 

Our observations confirm that  VY CMa has experienced several mass loss events, 
presumably from localized active regions. These clouds or loops of gas and dust are now
 expanding outward into what appears to be a surrounding nebula of surprisingly 
stationary gas and dust. This stationary or slow moving material material  
is difficult to explain around a star that has 
had a long history of high mass loss. We can only  speculate that the more diffuse, uniformly 
distributed gas has been created by a continuous, but slow general outflow of material 
from the overall surface. It therefore  seems  likely that the high mass loss for rate for VY CMa 
is really a measure of the mass carried out by these specific ejections  
perhaps accompanied by streams or flows of emitting gas through holes 
or gaps in the dust envelope.  
 
VY CMa is the only M--type  supergiant known to have a large, visible reflection nebula. 
Schuster \& Humphreys (2005)  have obtained HST/WFPC2 images  
of several cool hypergiant stars including the OH/IR M supergiants NML Cyg, S Per and VX Sgr 
to search for associated 
nebulosity. The results are basically negative or yield only marginal detections, except for 
NML Cyg which has a small 
bean-shaped nebula that has probably been shaped by the strong ionizing winds from the 
nearby hot stars in the Cygnus X complex \citep{MJ83}. 
In many ways VY CMa is most like IRC +10420, another high mass--loss, 
 strong OH/IR source believed to be in a post--red supergiant stage and with a
complex circumstellar environment \citep{TJJ,RMH97}. Its extended reflection 
nebula has a variety of structures
including knots and condensations, ray-like features and numerous arcs and scalloped loops which
in the outer parts closely resemble the arcs observed in VY CMa's ejecta. Spectroscopy
of its ejecta though showed a uniformly expanding outflow of gas \citep{RMH02}; there 
was no evidence for stationary gas or dust in its ejecta. 

The referee has suggested that the slow moving gas is due to 
gravitationally bound material in circular orbits (Jura and Kahane 1999).
This may be the case but the examples of orbiting material about evolved stars 
come from a very different class of objects; carbon-rich red giants and 
post AGB stars which are binaries or presumed binaries (Jura, Bahm, and Kahane 1995, 
Bujarrabel et al. 2003) with ``reservoirs'' of long -- lasting molecular material. 
At the distances where the motionless K I emission is observed (4000 - 10000 AU 
from the star), the orbital period will be several hundred thousand to a million
years, exceeding the expected duration of the red supergiant stage for a star this luminous. 
Furthermore, the OH and H$_{2}$O maser observations show that the clouds are expanding 
at 30 -- 40 km s$^{-1}$, greater than the expected escape velocity from the dust shell. 
Therefore we think this explanation is less likely for VY CMa if it is an evolved 
red supergiant.  

VY CMa is  a very perplexing star experiencing discrete mass loss
episodes that may be the primary mechanism by which it is losing so much mass. We have also
uncovered some puzzling  questions about the origin of its extended diffuse nebula with the slow
moving gas and dust. If it is an evolved massive star, it is apparently in a unique stage, 
with no obvious counterparts. Alternatively, the apparently ``stationary''  ejecta with 
embedded streams of gas and massive outflows is also consistent  
with a pre-main sequence state for VY CMa.

\acknowledgments
We are especially grateful to George Herbig for his continuing interest in VY CMa,
his comments and advice on this study, and for obtaining the telescope time for these
observations.  It is also a pleasure to thank Michael Koppelman for assistance with 
the data reduction.  This work was supported in part by the University of Minnesota.

Facilities:  \facility{Keck1(HIRES)}

\appendix

\section{Paradoxical K~I $\lambda$7699 emission}

The K~I emission line is difficult to assess, because this 
atom has an extremely small ionization potential 
(4.34 eV) and because some critical parameters are
inaccessible for VY CMa.  Since K~I ${\lambda}{\lambda}$7665,7699 
is a low-excitation resonance doublet,  
one is tempted to think ``it's merely resonance scattering''  
-- which, however, does not answer the most important questions.
Note, for instance, the strength of $\lambda$7699 emission 
relative to the continuum when we look directly toward the star, 
position I Ap 2 in Fig.\ 3.  In order to achieve the observed 
equivalent width of $\sim$ 5 {\AA} in emission simply by resonance 
scattering, a substantial amount of radiation must be removed from
the continuum along some (indeed  most) lines of sight. 
On average, {\it resonance scattering in itself neither augments 
nor diminishes the net emergent flux of $\lambda$7699 photons,\/}
excepting the small fraction that are scattered back into the
photosphere.  However, there is no hint of sufficient ``absorption''
to balance the observed ``emission'' anywhere along our slit positions, 
including reflected spectra that represent the star as seen from 
various directions in space.  Other puzzles involve the apparent 
velocity and the line profiles seen in the extended halo,
\S 4.2 above.

As a preliminary detail, note the optical thickness for 
K~I $\lambda$7699.  ($\lambda$7665 has a larger oscillator strength 
but is not included in our data.)  Assuming a normal potassium
abundance $n_K \; \approx \; 10^{-7} \, (n_H + 4n_{He})$ 
and a local line width of 4 km s$^{-1}$, the line-center opacity is 
${\kappa}_{\lambda7699} \; \sim \; 
10^{5} \, n(K^0) / n_K$ cm$^2$ g$^{-1}$.  
At least in the dusty zones, compare this to the continuum
opacity for scattering by dust grains,
${\kappa}_{d,sc} \; \sim \; 50$ cm$^2$ g$^{-1}$ at the same wavelength.
If K~I resonance scattering is to be at least comparable to
reflection by dust, we require  $n(K^0) \; > \; 5 \times 10^{-4} \, n_K $,
i.e., $n(K^+)/n(K^0) \, < \, 2000$.  
Given potassium's low ionization potential, this is a stronger 
constraint than one might guess at first sight.

\subsection{Potassium in the inner wind}

Can K~I emission arise within $r \, < \, 100$ AU, before dust has formed?
First consider the $K^+/K^0$ ionization ratio, which, in a simple model, 
depends on the star's photon supply at energies above 4.3 eV and on the 
electron density in the wind.  Suppose that free electrons are 
provided by the elements with low ionization potentials, 
giving $n_e \; \sim \; 6 \times 10^{-5} \, (n_H + 4n_{He})$.
At $r = 30$ AU, for example, in a simple model we expect 
$n_e \, \sim \, 10^5$ cm$^{-3}$.
If the continuum spectrum of VY CMa were a 3000 K Planckian, then it would 
produce about 10$^{46}$ potassium-ionizing photons per second;  but 
this is embarrassingly sensitive to temperature and other parameters, 
and the star's UV brightness may considerably exceed a simple 
black-body value (compare $\alpha$ Ori, Glassgold \& Huggins 1986).  Anyway, using 
reasonable estimates of the photoionization cross-section and 
recombination coefficient, the values quoted above imply 
   \begin{displaymath}
     n(K^+)/n(K^0) \; \sim \; 2 \times 10^4 ,  
   \end{displaymath}
which applies throughout the inner wind,  
12 AU $<$ $r$ $<$ 100 AU.  This ratio would be decreased by 
any additional supply of free electrons, e.g., by some 
process that can ionize a tiny fraction of the hydrogen and/or 
an appreciable fraction of carbon;  but it is difficult to 
reduce the K$^+$/K$^0$ ratio by a factor much smaller than 0.2 
in this way.  Most other likely complications tend to {\it increase\/} 
the ionization ratio, see Glassgold \& Huggins (1986).  Altogether, at this level 
of reasoning,  one expects the line-center opacity in the inner 
wind to be rather small for a resonance line, 
$\kappa_{\lambda7699} \, < \, 30$ cm$^2$ g$^{-1}$ .  
The total line-center optical depth along a radial path may 
exceed 10, but if so the effect on the total emergent spectrum 
is merely a fractional-{\AA} wavelength redistribution of photons 
near 7699 {\AA} -- i.e., quite different from the observed strong
emission feature which is not accompanied by a comparable amount 
of absorption at nearby wavelengths. 

If there are large-scale inhomogeneities in the local scattering
coefficient $ \rho \kappa_{\lambda7699} $, the emergent
spectrum can depend on viewing direction;  a net surplus of 
7699 {\AA} photons may escape in some directions at the cost of a 
net deficit elsewhere.  In order to obtain the observed strong feature 
in this way, however,  one must view the star from a 
statistically unusual direction.  A fundamentally non-spherical 
model in this vein -- for instance, imagine a circumstellar
torus seen from a near-axis viewpoint -- is difficult to construct
for the following reason:  In order to obtain the large observed
ratio ($\lambda$7699 peak height)/(continuum level) $\sim$ 5, the 
solid angle of ``resonance photon escape'' directions must be far 
smaller than that of the ``opaque'' directions.  (Admittedly this
may be the case south of the star, along Slit V.)  In summary, this 
general type of scenario cannot be rejected outright but we hope 
to find alternatives that are quantitatively more plausible.   

% paragraph break 

Can the observed K~I emission be excited by collisional processes 
in the inner wind?  In such a model the required luminosity 
in the $\lambda$7699 line is of the order of 150 $L_{\odot}$ or
$6 \times 10^{35}$ erg s$^{-1}$.  Electron collisions are rather
ineffective because $n_e$ is insufficient.  
If the inner-wind temperature is, say, 
4000 K, and we adopt the relative electron density mentioned above,
then collisional excitation of K~I $\lambda\lambda$7665,7699 proves
too weak by a factor of $\sim \; 10^{-3} \, n(K^0) / n_K $, i.e., 
by 5 to 9 orders of magnitude.  Radiative recombination is too weak
by about 8 orders of magnitude.   Unusual effects such as collisional 
excitation by fast non-thermal electrons, collisional excitation by 
hydrogen atoms, etc., are likewise unpromising, though 
we omit details here.  

Should we consider fluorescence, radiative excitation to higher 
levels of K~I? The upper level for $\lambda\lambda$7665,7699 is 
4p $^2$P$^{\rm o}$,  and the next resonance level, 5p $^2$P$^{\rm o}$, 
can be excited from the ground level by photons 
near 4044 {\AA}.  An ensuing cascade, 
   5p $\rightarrow$ (3d or 5s) $\rightarrow$ 4p, 
would lead to $\lambda$7699 emission.  Since the photosphere of 
VY CMa is relatively faint at violet wavelengths, this process seems  
worthwhile only if it is pumped by some other strong emission 
line near 4044 {\AA}.  So far as we know, there is no likely 
candidate for this role, nor for pumping K~I levels above 5p.

% paragraph break

In summary, we can almost eliminate all the obvious excitation 
processes other than resonance scattering, and even the latter
does not seem promising in the inner wind.   Therefore let us
focus on resonance scattering farther out, $r > 100$ AU,
where inhomogeneous dust may help to solve the puzzle.

\subsection{K~I $\lambda$7699 in zones where dust has recently formed}

Once dust has begun to form near $r_d \, \sim \, 100$ AU in the
outward mass flow, the K$^+$/K$^0$ ionization ratio becomes even
more difficult to predict.  The UV flux is greatly reduced by
even a small column density of dust, while, on the other hand,   
some ion species that help provide $n_e$ become depleted.  
Meanwhile the recombination time is several years, long enough 
for the gas to move a distance comparable to $r_d$.  Since the
data indicate that the K~I column density is indeed substantial
in dust farther out, let us {\it assume\/} that 
$\kappa_{\lambda7699} \, > \, 30$ cm$^2$ g$^{-1}$, 
i.e., at least comparable to scattering by dust, in the inner
parts of the dusty region.  This assumption is questionable but
it is not implausible.   

% paragraph break

Figure 14 shows, in an idealized way, how resonance scattering
can result in a net emission line if large-scale dusty condensations
are present.  Suppose, for the sake of illustration, that -- 
\\  % line break
$\bullet$ Each localized dense blob in the figure contains enough
dust to be practically opaque at all relevant wavelengths;  
\\  % line break
$\bullet$ Most photons from the star encounter such an obstacle;  
\\  % line break 
$\bullet$ Along some escape paths -- not necessarily straight --
the optical thickness of the lower-density medium is less than 3;   
\\  % line break
$\bullet$ The parameters allow about one or two percent of the 
red-wavelength continuum radiation to escape, as observed for VY CMa.  
Often the escape process involves an encounter with one of the dense 
blobs, scattering by a dust grain,  and perhaps further scattering 
in the less-dense medium.  The average dust-grain albedo is obviously 
relevant.  

In Fig.\ 14, a 7699 {\AA} photon from the star encounters a dusty 
condensation, is resonance-scattered by a potassium atom, and then escapes after
one more scattering event which may be due to either a K~I atom or
a dust grain.  This type of behavior, averaged over all photons,
leads to {\it an apparent emission line in the emergent spectrum.\/}  
There are at least two equivalent ways to see why:  
\\  % line break
(1) Temporarily imagine, for simplicity, that the dust absorbs but does 
not reflect radiation.  In that case, if the photon in Fig.\ 14 had not 
been scattered by K~I, then it would have been absorbed by dust in
the dense condensation..  This answers a question we asked above, 
namely, why does the observed spectrum not show K~I $\lambda$7699 
absorption lines to balance the 
apparent emission?  The answer is that most $\lambda$7699 scattering 
events occur in locations where the continuum radiation would not
otherwise escape anyway, being absorbed by dust if it isn't scattered 
by K~I.  
\\  % line break
(2) Of course, real dust grains reflect radiation as well as absorbing
it.  In effect, resonance scattering by K~I enhances the overall albedo
at a wavelength of 7699 {\AA}.  In other words, resonance scattering 
provides an extra way for 7699 {\AA} photons to escape from a dusty 
condensation.  Once they escape into a less-dense channel, they    
have an appreciable probability of escaping from the configuration.  
(Moreover, K~I scattering may be weak in the low-density regions,
so it does not necessarily impede the final escape process.)

The main point is that the K~I emission line may be ``formed'' by an
indirect process at radii beyond 100 AU or so.  Our observations 
indicate that expansion of the dusty medium does not red-shift 
the emergent K~I emission line as much as it does the stellar 
absorption spectrum (see \S 3);
but a quantitative model is necessary to judge whether this is
to be expected. Unfortunately, a model of this type has many adjustable 
parameters and probably requires Monte-Carlo-style radiative transfer
calculations, far beyond the scope of this paper.  

   % paragraph break

Similar reasoning can probably be applied to the Ca~I] emission.
However, as we noted in \S 4, the existence of Ca~I] in both 
emission and absorption seems to require an oscillator strength
far larger than one would expect for a semi-forbidden transition.
If this statement is wrong due to some geometrical or 
radiative-transfer fallacy, we have not been able to identify
the error.


%% copy editing.  Individual instruments can be provided in parentheses,
%% after the keyword, but they will not be verified.

\clearpage

\begin{figure}
\figurenum{1}
\epsscale{0.5}
\plotone{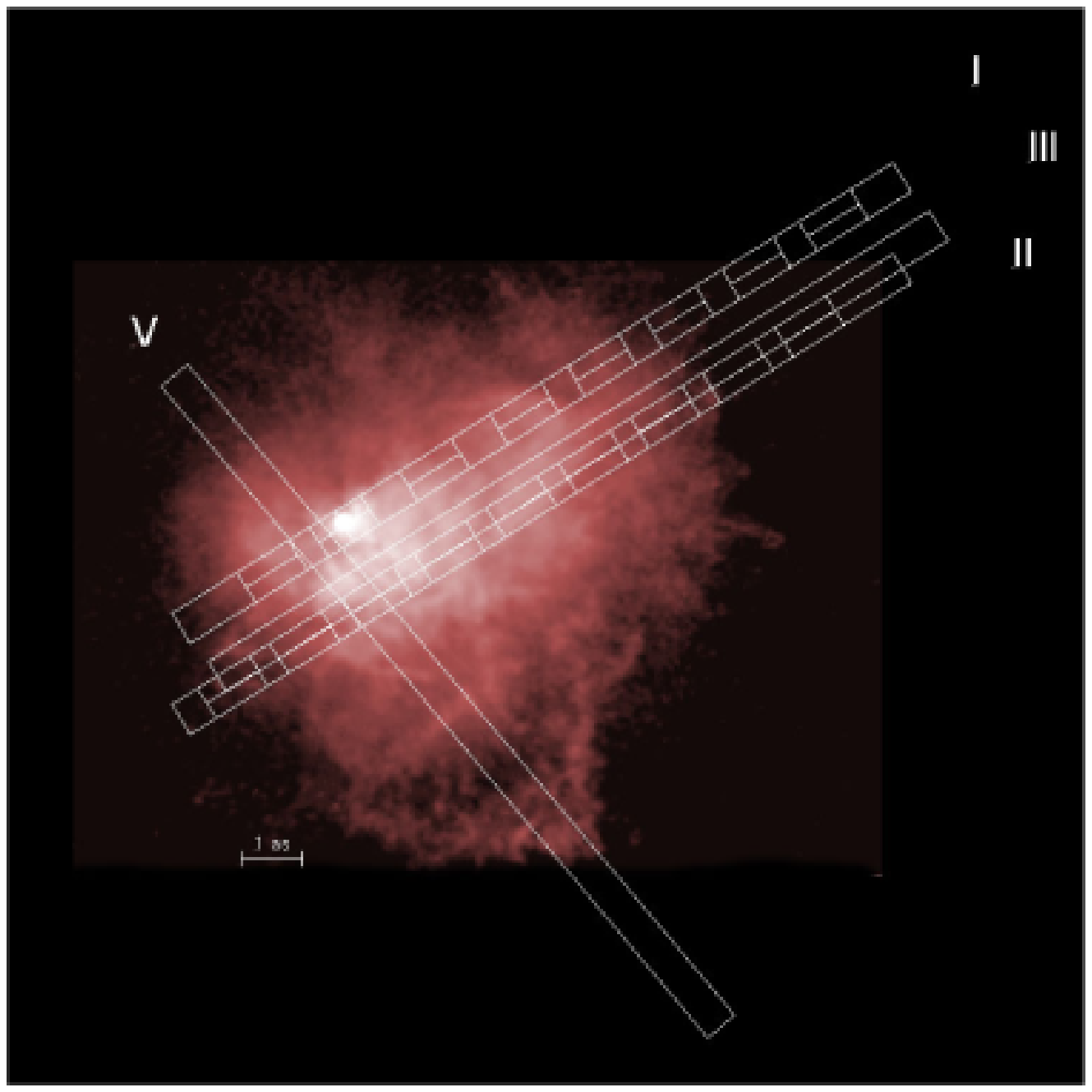}  
\caption{a. The HST/WFPC2 image showing the positions of the slits and the extraction apertures for Slits I and II.}  
\figurenum{1}  
\plotone{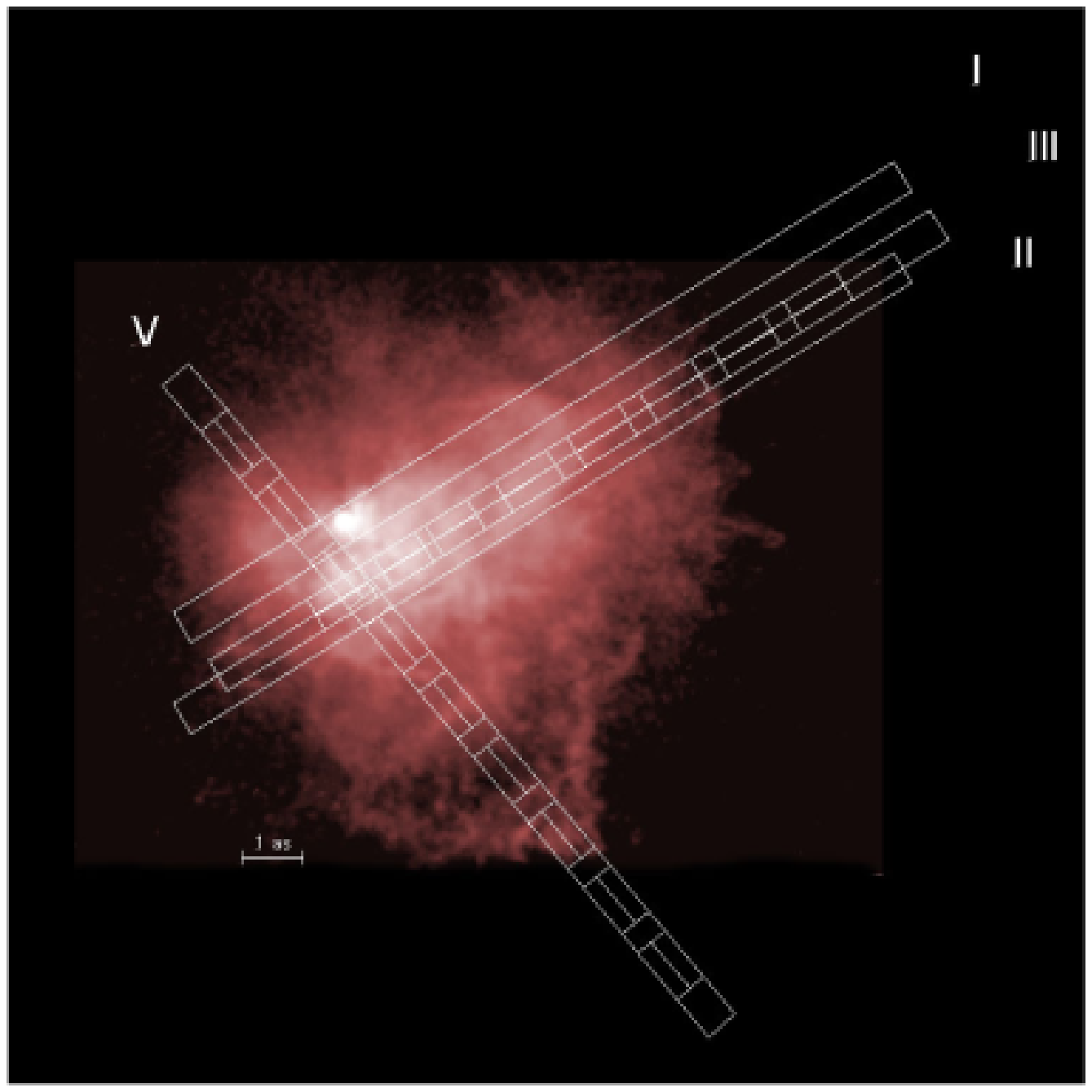}   
\caption{b. The HST/WFPC2 image showing the positions of the slits and the extraction apertures for Slits III and V.}
\end{figure}

\begin{figure}
\figurenum{2}
\epsscale{0.8}
\plotone{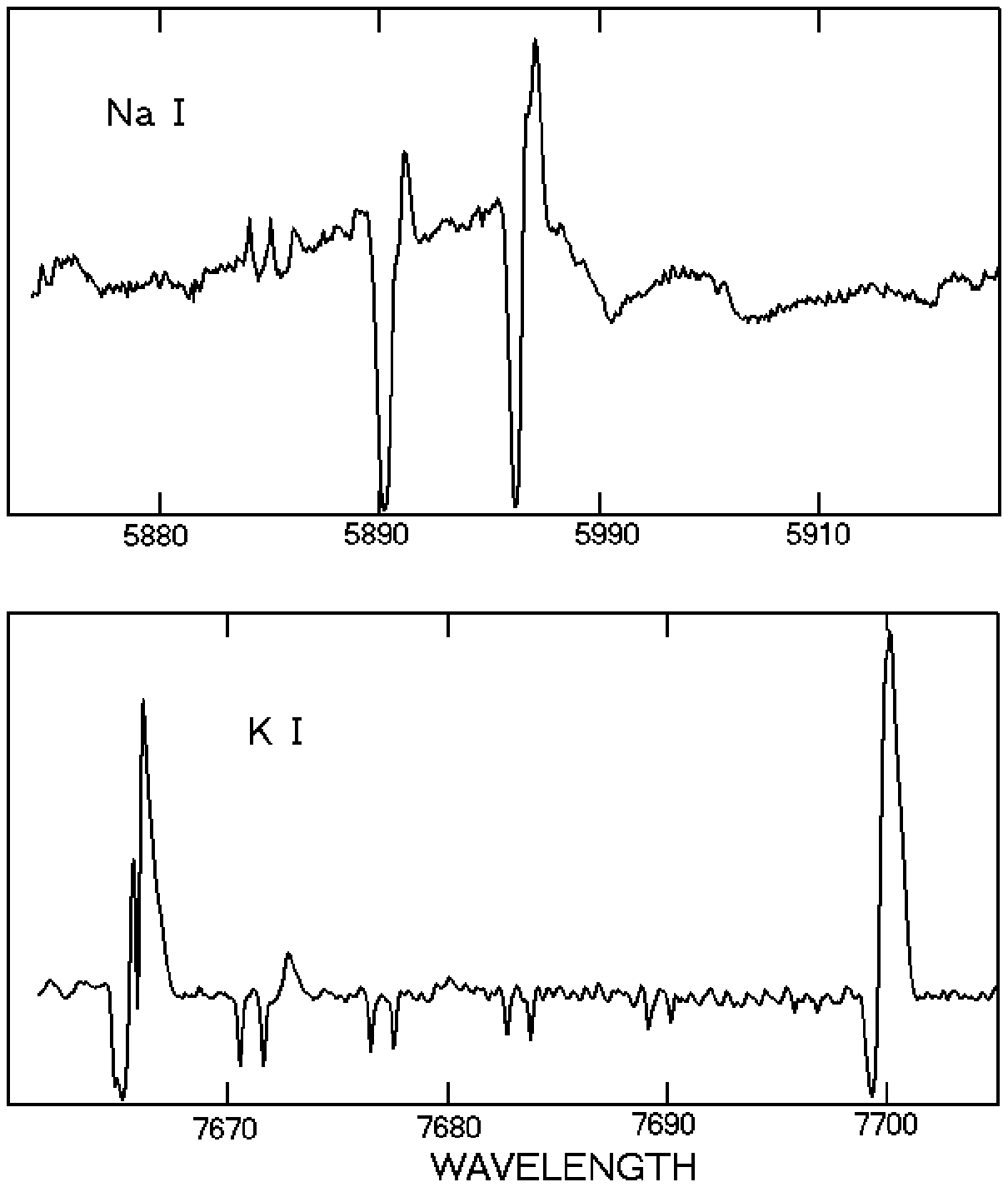}
\caption{Examples of the Na I and K I line profiles on the star from a spectrum obtained by G. Herbig with the same instrument.}
\end{figure}

\begin{figure}
\figurenum{3}
\epsscale{}
\plotone{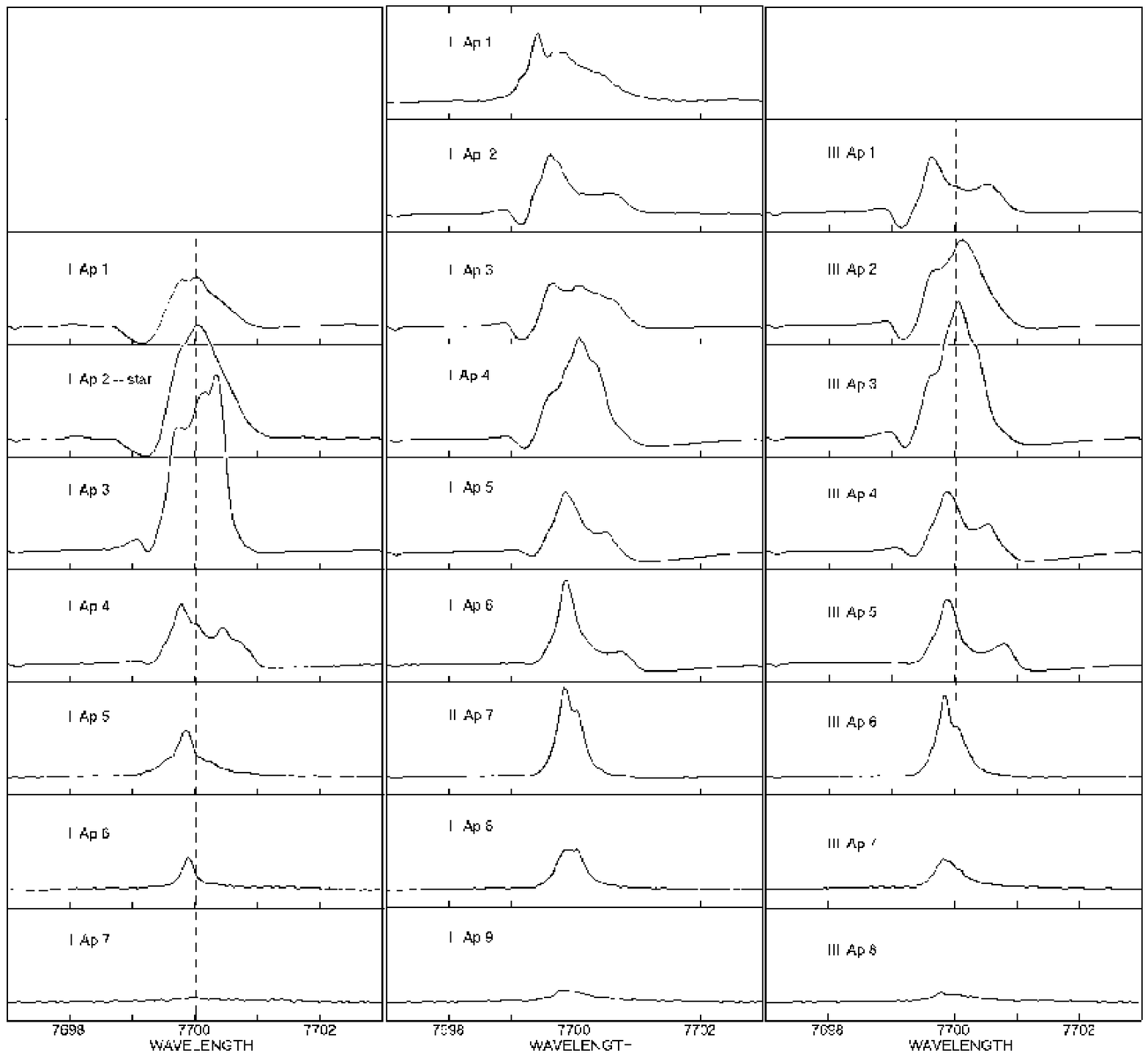}
\caption{The K I emission line profile along slits I, II and III. The dashed vertical line shows the velocity of the K I line at the star, Slit I, Ap.2.}
\end{figure}

\begin{figure}
\figurenum{4}
\epsscale{0.7}
\plotone{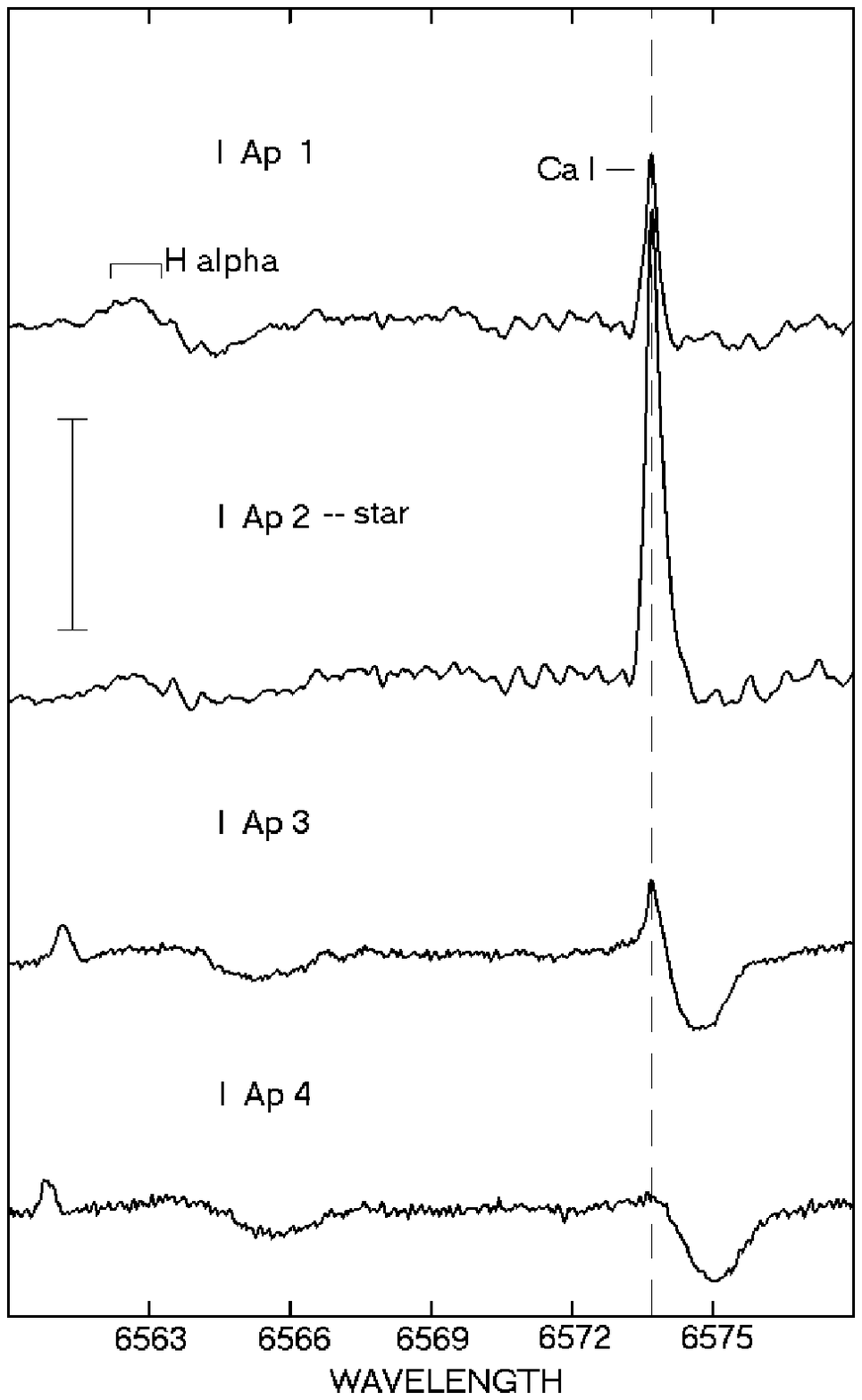}
\caption{The Ca I emission and absorption feature  and the H$\alpha$ region along Slit I. The dashed vertical line marks the velocity of the Ca I emission on the star. The Ca I emission velocity shows very little shift in all of the apertures. The narrow vertical line shows the height of the continuum.} 
\end{figure}

\begin{figure}
\figurenum{5}
\epsscale{0.8}
\plotone{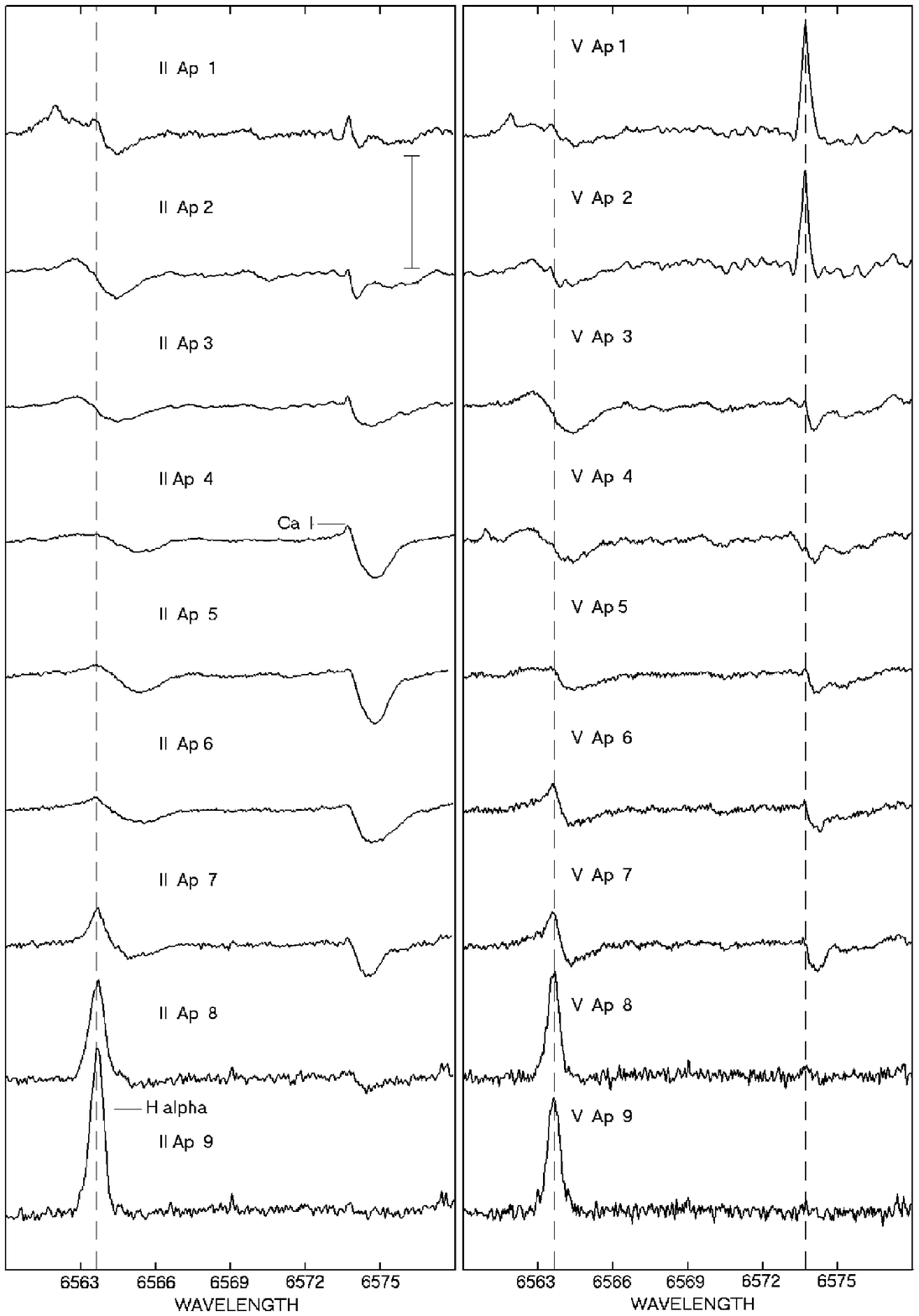}
\caption{The Ca I emission and absorption feature  and the H$\alpha$ region along Slits II and V. The dashed vertical line through H$\alpha$ marks a velocity of 37 km sec$^{-1}$, typical of its velocity in the outer parts of the nebula. The dashed vertical line  through  the Ca I emission line indicates its  velocity at the star. The narrow vertical line shows the height of the continuum.} 
\end{figure}

\begin{figure}
\figurenum{6}
\epsscale{0.9}
\plotone{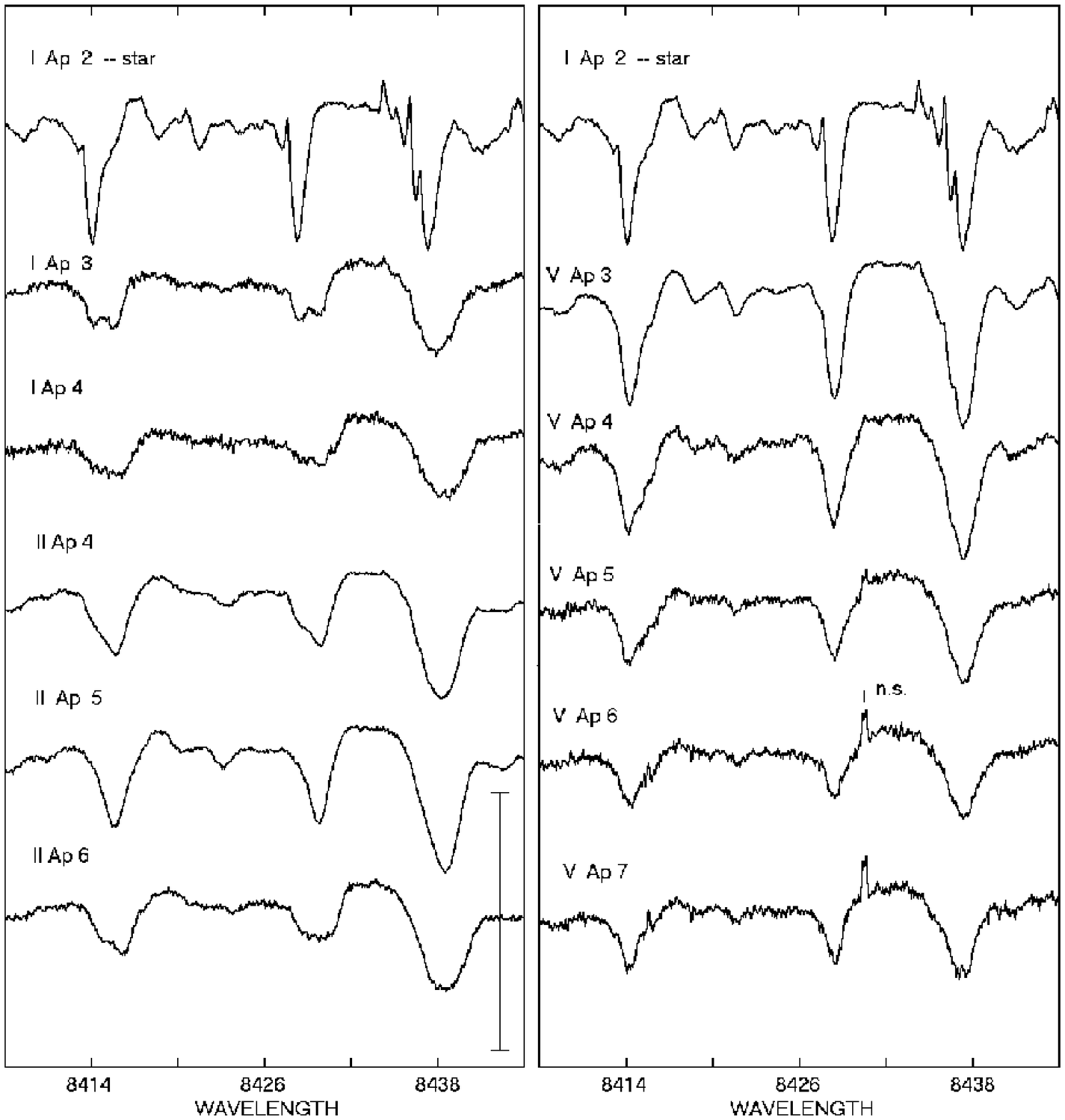}
\caption{Representative Fe I and Ti I absorption line profiles along Slits I, II and V showing the broadening of the absorption lines compared with the much narrower absorption lines on the star, Slit I Ap. 2. The narrow vertical line illustrates the height of the continuum.}  
\end{figure}

\begin{figure}
\figurenum{7}
\epsscale{1.10}
\plotone{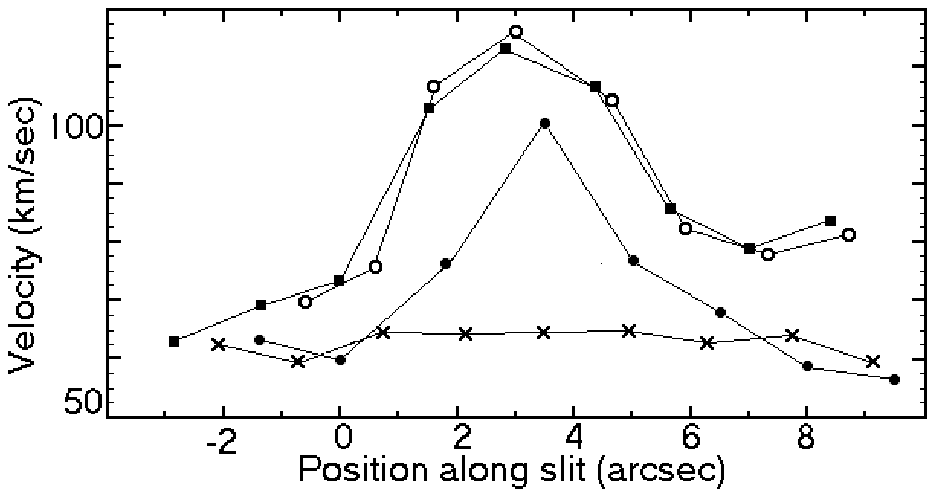}
\caption{The variation of the absorption line velocities with respect to the point nearest the star along Slits I ($\bullet$), Slit II ($\blacksquare$), Slit III ($\circ$) and Slit V ($\times$).}
\end{figure}

\clearpage

\begin{figure}
\figurenum{8}
\epsscale{0.5}
\plotone{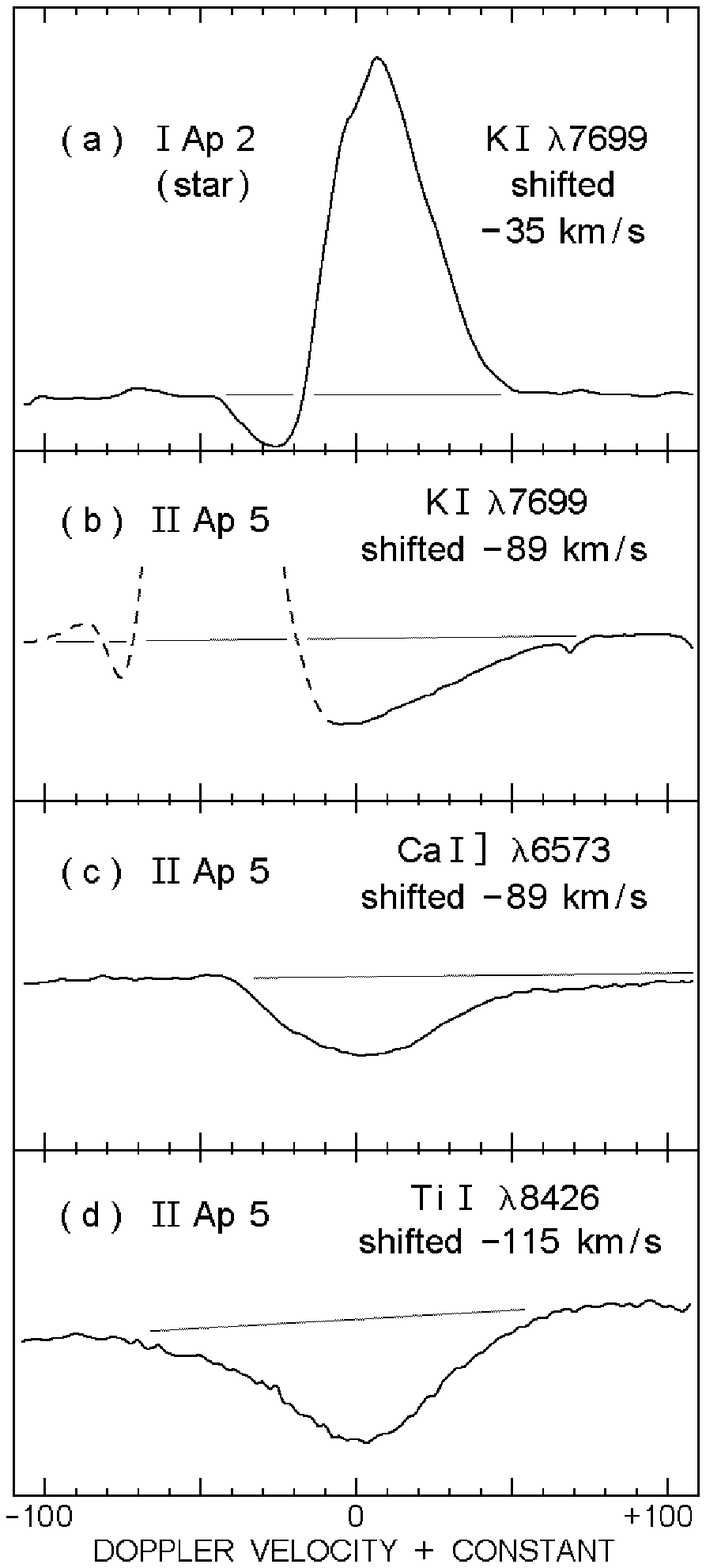}
\caption{Representative emission and absorption line profiles shifted to the rest frame of the system.The K I and Ca I] absorption lines are shifted by the expansion velocity plus the velocity of the system, the Ti I line by the above plus the additional redshift from the dusty shell, and the K I emission at the star by the system velocity only.}
\end{figure}

\begin{figure}
\figurenum{9}
\epsscale{0.9}
\plotone{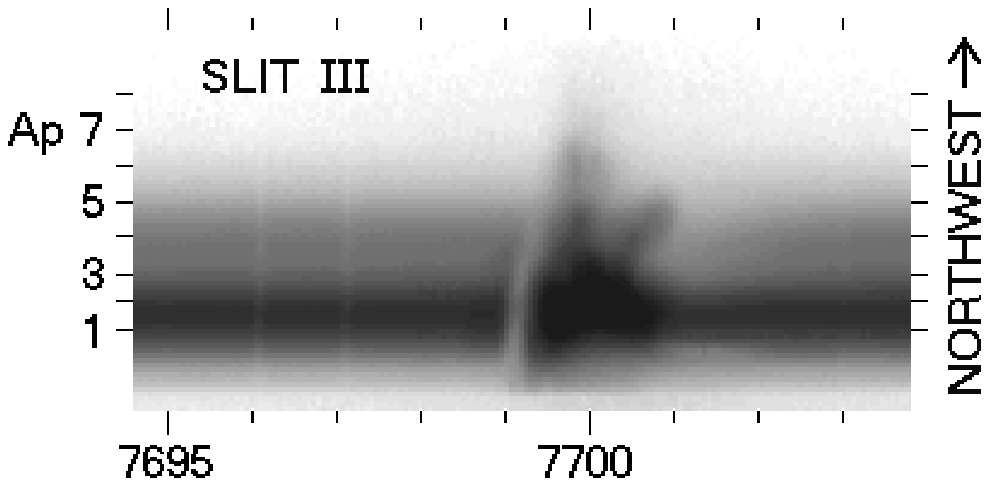}
\caption{The two-dimensional image of the K I emission line on Slit III. The straight emission feature on the blue side is the  constant velocity gas near the systemic velocity and the redward emission feature is probably due to gas moving outwards with the NW arc. The positions of the extraction apertures are marked.} 
\end{figure} 

%\clearpage 

\begin{figure}
\figurenum{10}
\epsscale{1.0}
\plotone{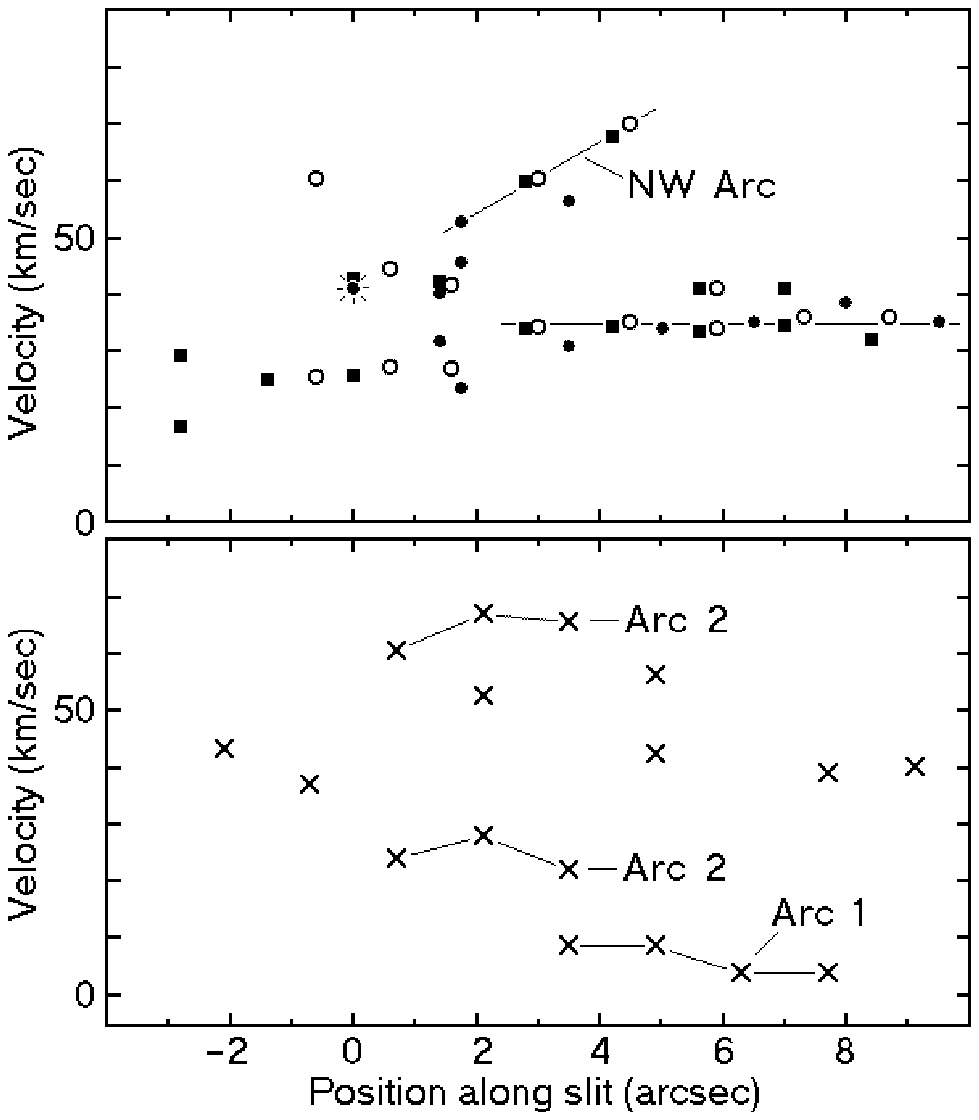}
\caption{The variation of the emission line velocities with respect to the point nearest the star along Slits I ($\bullet$), Slit II ($\blacksquare$), Slit III ($\circ$) and Slit V ($\times$). A line in the upper panel  connects the emission line velocities  across the NW arc and a horizontal line marks the systemic veocity of 35 km s$^{-1}$. In the lower panel for Slit V, the lines connect the velocities measured in the three ``streams'' of emitting gas in apertures 3 through 7.}
\end{figure}

\begin{figure}
\figurenum{11} 
\epsscale{0.5}
\plotone{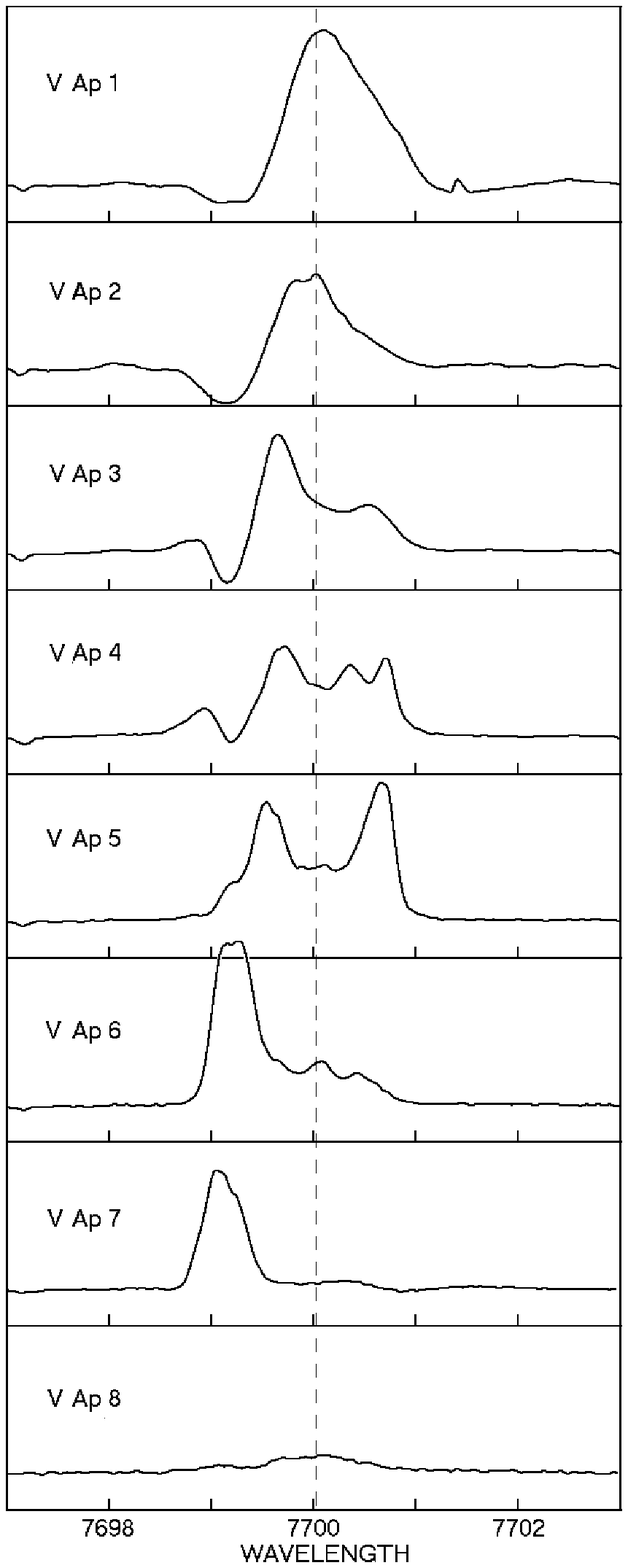}
\caption{The variation of the K I emission line profiles along Slit V. Note the  shift of the line peak to shorter wavelengths, apertures 5 to 7. The dashed vertical line marks the K I line velocity on the star.}
\end{figure}

\begin{figure}
\figurenum{12}
\epsscale{0.9}
\plotone{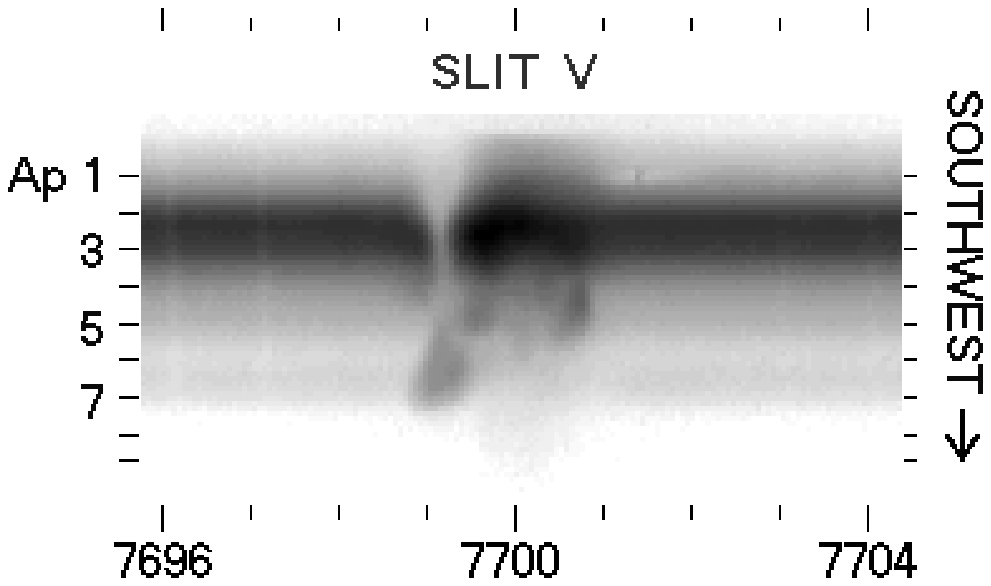}
\caption{The two-dimensional image of the K I emission line on Slit V. The positions of the extraction apertures are marked.}
\end{figure}

\begin{figure}
\figurenum{13}
\epsscale{0.8}
\plotone{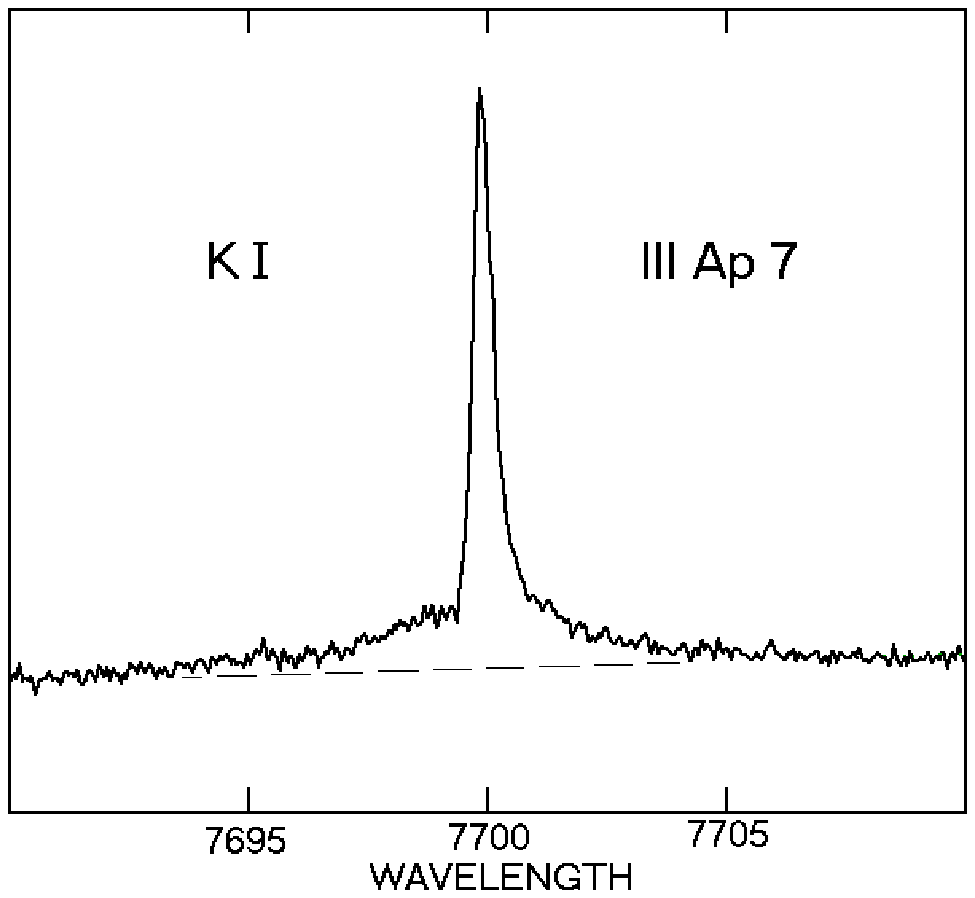}
\caption{A K I emision line from the outer parts of the nebula illustrating the narrow profile and the broad wings.}
\end{figure}

\begin{figure}
\figurenum{14}
\epsscale{0.6}
\plotone{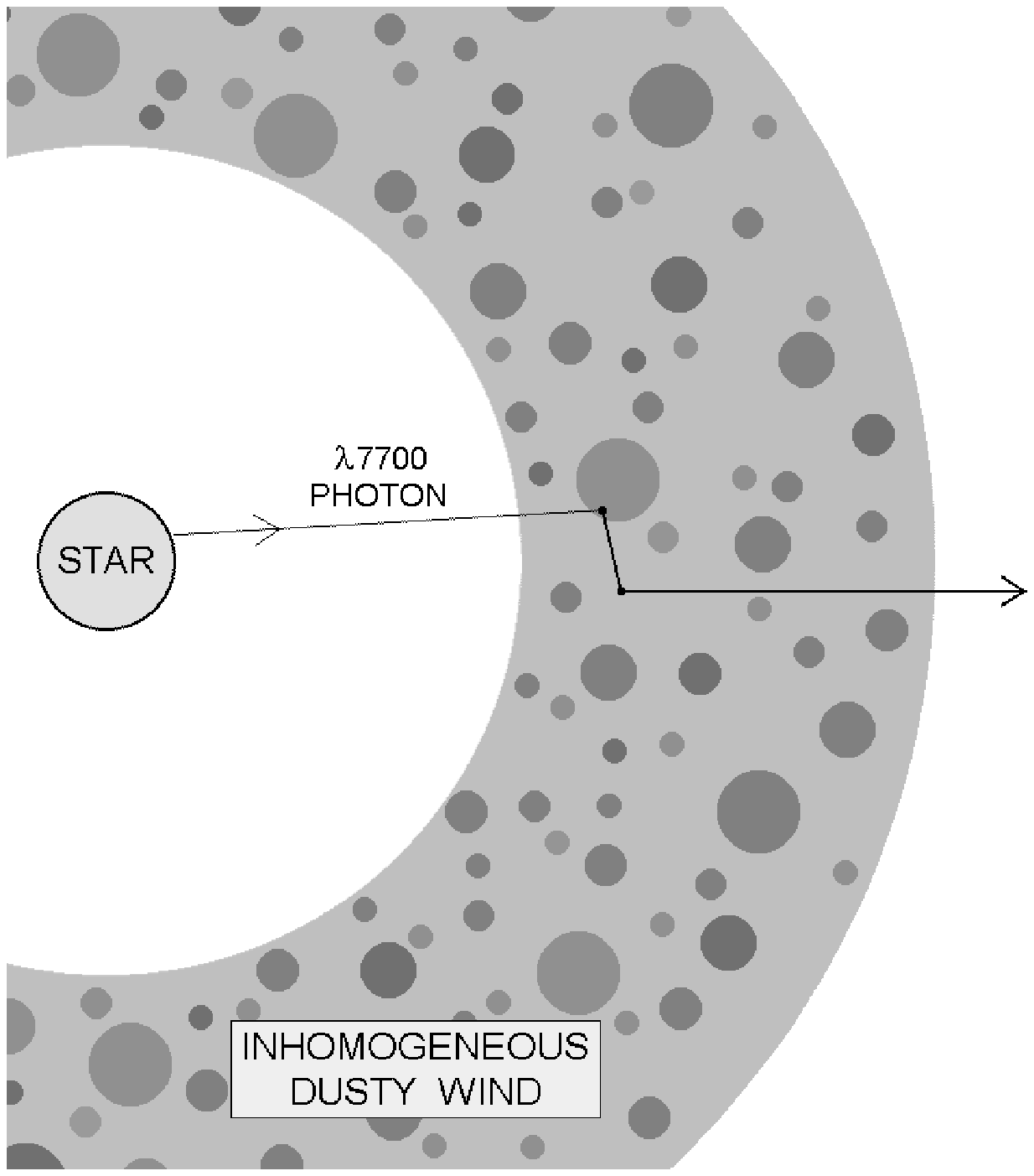}
\caption{A highly idealized illustration of how  scattering by large dusty condensations in the envelope can produce a resonance-scattered emission line.}
\end{figure}

\clearpage

%Table 1

\begin{deluxetable}{lc}
\tablecaption{Journal of Keck HIRES Observations, 15--16 December, 2002 }
\tablewidth{0pt}
\tablehead{
\colhead{Slit Position\tablenotemark{a}}  & \colhead{Exposure Times}
}
\startdata
I  &  60s, 120s \\
II &  4 $\times$ 600s \\
III & 60s, 3 $\times$ 600s\\
V   & 60s, 5 $\times$ 600s\\
\enddata
\scriptsize
\tablenotetext{a}{See Figure 1}
\end{deluxetable}

\clearpage 

%Table 2 
\begin{deluxetable}{lcccc}
\tablecaption{Extraction Positions Along Each Slit}
\tablewidth{0pt}
\tablehead{
\colhead{Slit} & \colhead{Aperture\tablenotemark{a}} & \colhead{Position Along Slit} & \colhead{Radial Distance} 
 & \colhead{Position Angle}\\
  &  & \colhead{from point nearest star} &  \colhead{from Star} & \colhead{from Star}\\
& & \colhead{in arc seconds} & \colhead{in arc seconds} & \colhead{in degrees} 
}
\startdata
I  &  1 &  1.40   & 1.40 & 122 \\
   &  2\tablenotemark{b} &  N.A.  & 0.00 &  0\\
   &  3 &  1.75  & 1.75 & -58\\
   &  4 &  3.50\tablenotemark{c}  & 3.50 & -58\\
   &  5 &  5.00  & 5.00 & -58\\
   &  6 &  6.50  & 6.50 & -58\\
   &  7 &  8.00  & 8.00 & -58\\
   &  8 &  9.50  & 9.50 & -58\\
II &  1 &  -2.80 & 3.23 & 145\\
   &  2 &  -1.40  & 2.02 & 161\\
   &  3 &   0.00  & 1.28 & -156\\
   &  4 &   1.40  & 1.77 & -104\\
   &  5 &   2.80\tablenotemark{c}  & 2.93 & -84\\
   &  6 &   4.20  & 4.23 & -76\\
   &  7 &   5.60  & 5.58 & -71\\
   &  8 &   7.00  & 6.95 & -69\\
   &  9 &   8.40  & 8.34 & -67\\ 
III & 1\tablenotemark{d} &  -0.60  & 1.21 & 178\\
    & 2\tablenotemark{e}\tablenotetext{f}  & 0.60 & 1.12 & -121\\
    & 3\tablenotemark{f} & 1.60 & 1.81 & -91\\
    & 4 &  3.00\tablenotemark{c}  &  3.13  & -76\\
    & 5 &  4.50  &  4.48  & -71  \\
    & 6 &  5.90  &  5.58  & -68  \\
    & 7 &  7.30  &  7.23  & -66  \\
    & 8 &  8.70  &  8.62  & -65  \\
V   & 1 & -2.10  &  2.28  & 54\\
    & 2 & -0.70  &  0.28  & 18\\
    & 3 &  0.70  &  0.81  & 177\\
    & 4 &  2.10   &  2.07  & -155\\
    & 5 &  3.50\tablenotemark{g}  &  3.44  & -149\\
    & 6 &  4.90  &  4.82  & -147\\
    & 7 &  6.30\tablenotemark{h}  &  6.22  & -145\\
    & 8 &  7.70  &  7.61  & -144\\   
    & 9 &  8.10  &  9.01  & -143\\
\enddata
\scriptsize
\tablenotetext{a}{Except where noted the apertures are 1$\farcs$0 $\times$ 0$\farcs$6.}
\tablenotetext{b}{The position of the star.}  
\tablenotetext{c}{NW arc}  
\tablenotetext{d}{S condensation}
\tablenotetext{e}{SW condensation}
\tablenotetext{f}{Aperture is 0$\farcs$8 $\times$  0$\farcs$6.}
\tablenotemark{g}{Arc 2}
\tablenotemark{h}{Arc 1}
\end{deluxetable}

\clearpage

%Table 3

\begin{deluxetable}{lccc}
\tablecaption{Measured Heliocentric Velocities of the Embedded Star in VY CMa}
\tablewidth{0pt}
\tablehead{ 
\colhead{Line} & \colhead{Velocity km s$^{-1}$}  & \colhead{Previous Velocities - Mean}
  & \colhead{and  Range}\\ 
& \colhead{km s$^{-1}$}  & \colhead{km s$^{-1}$}  & \colhead{km s$^{-1}$} 
}   
\startdata
K I emission  &  41.0    &  51.0  &   40 -- 58              \\
K I P Cyg absorption  & 7.8  & 6 & 5 -- 7                    \\
Ca I emission  &   41.9  &   49.2  &  36 -- 51       \\
Rb I emission      & 44.0   & \nodata     &  \nodata    \\
Absorption lines (9)\tablenotemark{a}  &  59.9  &  67.5  & 37 -- 83  \\
\enddata
\scriptsize
\tablenotetext{a}{Ca II, Fe I, Ti I}
\tablerefs{Wallerstein 1958, Hyland et al. 1969, Wallerstein 1971, Humphreys 1970, Humphreys 1975, Wallerstein 1977, Wallerstein \& Gonzalez 2001}
\end{deluxetable} 

\clearpage

%Table 4

\begin{deluxetable}{lcccccc}
\tablecaption{Emission Line Velocities Along Slits I, II, III}
\tablewidth{0pt}
\tablecolumns{7}
\tablehead{
\colhead{Aperture} & \colhead{Slit I em} & \colhead{P Cyg Abs.} & \colhead{Slit II em} 
& \colhead{P Cyg abs.} & \colhead{Slit III em} & \colhead{P Cyg abs}\\ 
&  \colhead{km s$^{-1}$} &  \colhead{km s$^{-1}$} &  \colhead{km s$^{-1}$} &  \colhead{
km s$^{-1}$} &  \colhead{km s$^{-1}$} &  \colhead{km s$^{-1}$}
}
\startdata
\sidehead{ K I emission and P Cyg absorption}  
1 & 31.6, 40.2 & 6.7 & 16.8, 29.3 & \nodata & 25.4, 60.4 & 6.3\\
2 & 41.0 & 7.8 & 25.0 & 6.3 & 27.3, 44.5 & 8.2\\
3 & 23.4, 45.6, 52.6 & 10.2 & 25.8, 42.5 & 6.3 & 26.9, 41.7 & 8.2\\
4 & 30.8, 56.5 & 10.2 & 42.1 & 8.2 & 34.3, 60.4 & 13.3\\ 
5 & 33.9 & \nodata & 33.9, 59.7 & 12.9 & 35.0, 70.0 & 12.0 \\
6 & 35.1 & \nodata & 34.3, 67.8 & 12.5 & 34.0, 41.0 & \nodata \\
7 & 38.6 & \nodata & 33.5, 41.0 & \nodata & 36.0 & \nodata\\
8 & 35.1 & \nodata & 34.7, 41.0 & \nodata & 32, 39 & \nodata\\
9 & \nodata   & \nodata &  32.0      & \nodata & \nodata & \nodata\\
\sidehead{ Ca I emission and P Cyg absorption}
1 & 41.5 & \nodata & 42.9 & 24.6 & 13.2, 40.1 & \nodata\\
2 & 41.9 & \nodata  & 41.9 & 31.0 &   39.7 & \nodata\\  
3 & 41.0 & \nodata  & 41.0 & \nodata   &   41.0 & \nodata\\
4 &  \nodata  & \nodata & 40.6 & \nodata  &  38.8 & \nodata\\
5 &  \nodata  & \nodata & 41.5 & \nodata   &   39   & \nodata\\
6 &  \nodata  & \nodata & \nodata  & \nodata   &   41   & \nodata\\
7 &  \nodata  & \nodata & \nodata  & \nodata   &   42    & \nodata\\
\tablebreak
\sidehead{ H$\alpha$ emission}
1 & -14\tablenotemark{a}     & \nodata & 34.0 & \nodata   &   -8.1 & \nodata\\
2 & \tablenotemark{b}     & \nodata & -7.6 & \nodata   &   -5.3 & \nodata\\
3 & 10 \tablenotemark{c}     & \nodata & -4.4 & \nodata   &   \nodata   & \nodata\\
4 &      & \nodata &  8.4 & \nodata   &  34.0 & \nodata\\
5 & \nodata   & \nodata & 35.3 & \nodata   & 36   & \nodata\\
6 & 37   & \nodata & 34.0 & \nodata   &  36    & \nodata\\
7 & 37   & \nodata & 37.2 & \nodata   &  36    & \nodata\\
8 & 37   & \nodata & 38.5 & \nodata   &  36   & \nodata\\
9 & \nodata   & \nodata & 37.6 & \nodata   &  \nodata    & \nodata\\
\enddata
\scriptsize
\tablenotetext{a}{broad emission}
\tablenotetext{b}{very weak, broad emission}
\tablenotetext{c}{broad emission}
\end{deluxetable}

\clearpage

%Table5

\begin{deluxetable}{lcccccccc}
\tablecaption{Absorption Line Velocities Along Slits I, II, III, and V}
\tablewidth{0pt}
\tablecolumns{9}
\tablehead{
\colhead{Aperture} & \colhead{Slit I} & \colhead{n lines} & \colhead{Slit II}  & \colhead{n lines}  & \colhead{Slit III}  & \colhead{n lines} &  \colhead{Slit V}  & \colhead{n lines}\\
& \colhead{km s$^{-1}$} & & \colhead{km s$^{-1}$} & & \colhead{km s$^{-1}$} & & 
\colhead{km s$^{-1}$} & 
}
\startdata
\sidehead{Ca II, Fe I, Ti I (8400 -- 8800$\AA$)}  
1 & 63.2 & 9 & 63.2 & 9 & 69.8 & 8 & 62.6 & 9\\
2 & 59.9 & 9 & 69.0 & 8 & 75.9 & 8 & 59.3 & 9\\
3 & 76.7\tablenotemark{a} & 8 & 73.6 & 8 & 107.9\tablenotemark{b} & 5 & 64.7 & 9\\
4 & 100.4 & 8 & 101.8\tablenotemark{c} & 4 & 116.0 & 8  & 64.4 & 8\\
5 & 76.7  & 5 & 113.6 & 8 & 102.3\tablenotemark{d}  & 7 & 64.6 & 8\\
6 & 68.1  & 3 & 107.4\tablenotemark{e} & 6 & 82.4  & 8 & 64.8 & 8\\
7 & 58.5  & 2 & 85.9  & 7 & 78.0  & 6 & 62.7 & 8\\
8 & 56.5  & 1 & 78.9  & 6 & 81.3  & 3  & 64.2 & 4\\
9 & \nodata  & \nodata & 83.9 & 3 & \nodata  & \nodata  & 59.5 & 3\\
\sidehead{Ca I }  
1 & \tablenotemark{f}  & & 63.4 & & 57.5 & & \nodata  &\\
2 & \tablenotemark{g} & & 58.8 & & 85.3 & & 64.8 &\\ 
3 & 89.9 & & 84.4 & & 93.1 & & 57.9 &\\
4 & 103.1 & & 92.6 & & 92.6 & & 59.7 &\\
5 & 80.8  & & 91.2 & & 99.0     & & 62.9 &\\
6 & \nodata    & & 89.0 & & 81.0     & & 68.9 &\\
7 & \nodata   & & 79.8 & & \nodata     & & 64.8 &\\
8 & \nodata    & & 76.2 & & \nodata     & & \nodata   &\\
\tablebreak
\sidehead{ K I}
1 &  \nodata   & & \nodata  & & \nodata   & & 92.0  \\   
2 &  \nodata   & & \nodata  & & 88.1   & &  \nodata  \\   
3 &   \nodata  & & 89.7     & & 86.5   & &  \nodata    \\   
4 &  86.9      & & 89.7     & & 84.6   & &  \nodata    \\   
5 &   \nodata  & & 84.6  & &   87.0    & &  \nodata     \\   
6 &   \nodata  & & 84.6  & &  \nodata  & &   \nodata   \\  
\sidehead{ H$\alpha$ }  
1 &48, 74 & & 76  & &  73.7 & & 92.0 &\\
2 & 48 & & 73 & &  86.5 & & 71.9 &\\
3 & 114 & & 76  & &  121.7 & & 73.7 &\\
4 & 132 & & 118 & &  118.1 & & 68.2 &\\
5 & \nodata& & 117  & &  126  & & 79.7 &\\
6 & \nodata & & 119  & & 109  & & 77.8 &\\
7 & \nodata & & 111 & & \nodata  & & 68.7 &\\
\enddata
\scriptsize
\tablenotetext{a}{Four of the lines are double. 76.7 km s$^{-1}$ is the mean velocity of the four single lines and the blues-hifted component of the double lines. The red-shifted component has a mean velocity of 110.1 km s$^{-1}$}
\tablenotetext{b}{Several of the lines in the aperture are obviously asymmetric with a hint of a blue-shifted second component. We were able to measure a velocity of 75.6 km s$^{-1}$ for this feature in three of the lines. The mean velocity given here is for the single lines and the absorption minimum of the asymmetric lines.}
\tablenotetext{c}{A few lines are asymmetric with a slight extension to the blue. The mean velocity is for the single lines. }
\tablenotetext{d}{Three of the lines are asymmetric to the red. The mean velocity is for the single lines plus the absorption minimum of the asymmetric lines.}
\tablenotetext{e}{Two of the lines are asymmetric to the red. The mean velocity is for the single lines.}
\tablenotetext{f}{no absorption apparent}
\tablenotetext{g}{absorption may be marginally present; partially covered by emission}
\end{deluxetable}

\clearpage

%Table6

\begin{deluxetable}{lccccc}
\tablecaption{Emission Line Velocities Along Slit V}
\tablewidth{0pt}
\tablehead{
\colhead{Aperture} & \colhead{K I} & \colhead{P Cyg} & \colhead{Ca I} & \colhead{P Cyg} & \colhead{H$\alpha$}\\
& \colhead{km s$^{-1}$} & \colhead{km s$^{-1}$} & \colhead{km s$^{-1}$} & \colhead{km s$^{-1}$} & \colhead{km s$^{-1}$}
}
\startdata
1 & 43.3 & 4.3 & 41.9 & 20.0 & -42.8, -2.1, 31.7\\
2 & 37.1 & 6.3 & 41.9 & 20.5 &        -3.5, 30.3\\
3 & 24.1, 60.4 & 6.7 & 40.6 & \nodata & -3.5\\
4 & 28.1, 52.6, 67.1 & 7.8 & \nodata & \nodata &  -14.5\\
5 & (8.6)\tablenotemark{a},21.9, 65.5 & \nodata & 41.5 &\nodata & \nodata\\
6 & 8.6, 42.5, 56.2 & \nodata & 40.6 & \nodata & 35.3\\
7 & 3.9 & \nodata & 38.3 & \nodata & 34.9\\
8 & (3.9)\tablenotemark{b}, 39.0 & \nodata & \nodata & \nodata & 35.8\\
9 &      40.2 & \nodata & \nodata & \nodata  & 37.6\\
\enddata
\scriptsize
\tablenotetext{a}{Velocity of the small bump on the blue side of the line profile.}
\tablenotetext{b}{A rather broad feature is weakly present at this velocity for its
``peak'' emission.}
\end{deluxetable}


\begin{thebibliography}{}
\bibitem[Bernat \& Lambert 1976]{BL76}Bernat, A. P. \& Lambert, D. L. 1976, \apj, 210, 395
\bibitem[Bowers, Johnston, \& Spencer 1983]{Bow83}Bowers, P. F., Johnston, K. J., \& Spencer, J. H. 1983, \apj, 274, 733
\bibitem[Bujarrabel et al. 2003]{Buj03}Bujarrabel, V., Neri, R., Alcolea, J. \& Kahane, C. 2003, \aap, 409, 573
\bibitem[Danchi et al. 1994]{Dan94}Danchi, W.C., Bester, M., Degiacomi, C.G., Greenhill, L.J., \& Townes, C.H.  1994, \aj, 107, 1469
\bibitem[de Jager 1998]{deJ98}de Jager, C. 1998, \aapr, 8, 145
\bibitem[Dupree, Lobel \& Gilliland 1999]{Dup99}Dupree, A.K., Lobel, A. \& Gilliland, R.L. 1999, \baas, 194, 6605
\bibitem[Efstathiou \& Rowan-Robinson]{Efs90}Efstathiou, A. \& Rowan-Robinson, M. 1990, \mnras, 245, 275.
\bibitem[Gilliland \& Dupree 1996]{GD96}Gilliland, R.L. \& Dupree, A.K. 1996, \apj, 463, L29
\bibitem[Glassgold \& Huggins 1986]{Gla86}Glassgold, A. E. \& Huggins, P. J. 1986, \apj, 306, 605 
\bibitem[Guilain \& Mauron 1996]{GM96}Guilain, G. \& Mauron, N. 1996, \aap, 314, 585  
\bibitem[Herbig 1970a]{GH70a}Herbig, G.H. 1970a, Mem. Soc. Roy. Liege, 19, 13
\bibitem[Herbig 1970b]{GH70b}Herbig, G.H.  1970b, \apj, 162, 557
\bibitem[Herbig 1972]{GH72}Herbig, G.H.  1972, \apj, 172, 375
\bibitem[Herbig 1974]{Her74}Herbig, G.H.  1974, \apj, 188, 533 
\bibitem[Humphreys 1970]{RMH70}Humphreys, R. M. 1970, \pasp, 82, 1158  
\bibitem[Humphreys 1975]{RMH75}Humphreys, R. M. 1975, \pasp, 87, 433
\bibitem[Humphreys \& Davidson 1994]{HD94}Humphreys, R.M. \& Davidson, K. 1994, \pasp, 106, 1025
\bibitem[Humphreys et al. 1997]{RMH97}Humphreys, R.M., et al. 1997, \aj, 114, 2778
\bibitem[Humphreys et al. 2002]{RMH02}Humphreys, R.M., Davidson, K., \& Smith, N. 2002, \aj, 124, 1026
\bibitem[Hyland et al. 1969]{Hy69}Hyland, A.R., Becklin, E.E., Neugebauer, G. \& Wallerstein, G. 1969, \apj, 15 8, 619
\bibitem[Jones et al 1993]{TJJ}Jones, T. J. et al. 1993, \apj, 411, 323
\bibitem[Jura \& Kahane 1999]{JK99}Jura, M. \& Kahane, C. 1999, \apj, 521, 302
\bibitem[Jura, Balm, \& Kahane 1995]{JBK95}Jura, M., Balm, S. P. \& Kahane, C. 1995, \apj, 453, 721 
\bibitem[Kluckers et al. 1997]{Kl97}Kluckers, V.A., Edmunds, M.G., Morris, R.H. \& Wooder, N. 1997, \mnras, 284, 711
\bibitem[Lada \& Reid 1978]{LR78}Lada, C.J. \& Reid, M.J.  1978, \apj, 219, 95
\bibitem[Lobel \& Dupree 2000]{LD00}Lobel, A. \& Dupree, A. K. 2000, \apj, 545, 454
\bibitem[Lobel \& Dupree 2001]{LD01}Lobel, A. \& Dupree, A. K. 2001, \apj, 558, 815
\bibitem[Marvel 1997]{Mar97}Marvel, K.B.  1997, \pasp, 109, 1286
\bibitem[Monnier et al. 1999]{Monn99}Monnier, J.D., Tuthill, P.G., Lopez, B., Cruzalebes, P., Danchi, W.C., \& Haniff, C.A.  1999, \apj, 512, 351
\bibitem[Morris \& Bowers 1980]{MB80}Morris, M. \& Bowers, P. F., 1980, \aj, 85, 724
\bibitem[Morris \& Jura 1983]{MJ83}Morris, M. \& Jura, M. 1983, \apj, 267, 179 
\bibitem[Nercessian et al. 1989]{Ner89}Nercessian, E., Guilloteau, S., Omont, A. \& Benayoun, J. J. 1989, \aap, 210, 225 
\bibitem[Richards, Yates \& Cohen 1998]{RYC98}Richards, A. M. S., Yates, J. A., \& Cohen, R. J. 1998, \mnras, 299, 319
\bibitem[Robinson 1971]{Rob71}Robinson, L. J. 1971, IBVS, 599
\bibitem[Schaller et al. 1992]{Sch92}Schaller, G., Schaerer, D., Meynet, G.  \& Maeder, A. 1992, \aaps, 96, 269 
\bibitem[Schaerer et al. 1993]{Sch93}Schaerer, D., Meynet, G., Maeder, A.  \& Schaller, G. 1993, \aaps, 98, 523
\bibitem[Schuster \& Humphreys 2005]{SH05}Schuster, M. T. \& Humphreys, R. M. 2005, in preparation
\bibitem[Smith 2004]{Smi04}Smith, N. 2004, \mnras,  
\bibitem[Smith et al. 2001]{Smi01}Smith, N., Humphreys, R. M., Davidson, K., Gehrz, R.  D., and Schuster, M. T. \& Krautter, J.  2001, \aj, 121, 1111 
\bibitem[Uitenbroek, Dupree \& Gilliland 1998]{Uit98}Uitenbroek, H., Dupree, A.K., \& Gilliland, R.L., 1998, \aj, 116, 2501
\bibitem[Wallerstein 1958]{Wall58}Wallerstein, G. 1958, \pasp, 70, 479
\bibitem[Wallerstein 1971]{Wall71}Wallerstein, G. 1971, \apj, 169, 195 
\bibitem[Wallerstein 1977]{GW77}Wallerstein, G. 1977, \apj, 211, 170
\bibitem[Wallerstein 1978]{Wall78}Wallerstein, G. 1978, Observatory, 98, 224
\bibitem[Wallerstein 1986]{Wall86}Wallerstein, G. 1986, \aaps, 164, 101  
\bibitem[Wallerstein 2001]{Wall01} Wallerstein, G. \& Gonzalez, G.  2001, \pasp, 113, 954 
\bibitem[Wittkowski et al. 1998]{Witt98} Wittkowski, M., Langer, N. \& Weigelt, G. 1998, \aap., 340, L39
\end{thebibliography}
\end{document}